\documentclass[14pt]{extarticle}

\usepackage[T1]{fontenc}
\usepackage[utf8]{inputenc}
\usepackage{graphicx}
\usepackage{dcolumn}
\usepackage{bm}
\usepackage{color}
\usepackage{soul}

\usepackage[scr=boondoxo,scrscaled=1.05]{mathalfa}

\usepackage[section]{placeins} 
\usepackage{amsmath}
\usepackage{amsfonts}
\usepackage{amssymb}
\usepackage{makeidx}
\usepackage{notes2bib}

\usepackage{xcolor}
\usepackage{mathtools}

\usepackage[a4paper,total={7in, 10in}]{geometry}

\newcommand{\nc}{\newcommand}
\nc{\nn}{\nonumber}
\def\e{\mathcal{E}}

\def\beq{\begin{equation}}
\def\eeq{\end{equation}}
\def\beqa{\begin{eqnarray}}
\def\eeqa{\end{eqnarray}}

\usepackage[normalem]{ulem}

\author{Y. Sivan$^{1,\ast}$, I. W. Un$^1$, I. Kalyan$^1$, K.-Q. Lin$^2$, J. M. Lupton$^2$, S. Bange$^2$ \\
\normalsize{$^1$ School of Electrical and Computer Engineering, Ben-Gurion University
, Israel}\\
\normalsize{$^2$ Institut für Experimentelle und Angewandte Physik, Universität Regensburg, 
Germany}\\
\normalsize{$^\ast$ To whom correspondence should be addressed; E-mail: sivanyon@bgu.ac.il} }

\title{
Crossover from non-thermal to thermal photoluminescence from metals excited by ultrashort light pulses}
\date{\today}

\begin{document}
\maketitle

\begin{abstract}
Photoluminescence from metal nanostructures following intense ultrashort illumination is a fundamental aspect of light-matter interactions. Surprisingly, many of its basic characteristics are under ongoing debate. Here, we resolve many of these debates by providing a comprehensive theoretical framework that describes this phenomenon, and support it by experimental confirmation. Specifically, we identify aspects of the emission that are characteristic to either non-thermal or thermal emission, in particular, differences in the spectral and electric field-dependence of these two contributions to the emission. Overall, non-thermal emission is characteristic of the early stages of light emission, while the later stages show thermal characteristics. The former dominate only for moderately high illumination intensities for which the electron temperature reached after thermalization remains close to room temperature. The theory is complemented by experimental evidence that demonstrates the novel aspects of our considerations.
\end{abstract}

\section{Introduction}

\subsection{General background}
The emission of light from metals following illumination, known as metal photoluminescence (PL), is a fundamental aspect of light-metal interactions. Surprisingly, this phenomenon is far from being understood, with its main features having been under debate for decades~\cite{Baffou-aSE-thermometry-review}. For example, there is not even agreement on whether the emission is due to electronic Raman scattering (whereby absorption and emission occur with no delay and with phase correlations) or due to radiative recombination (for which a delay does occur and the emission has no phase correlation with the exciting photon), see, e.g.,~\cite{metal_luminescence_Cahill_PNAS}. For the relatively simple scenario of emission under CW illumination, the emission statistics (i.e., bosonic or fermionic) or whether it is thermal or non-thermal (sometimes dubbed as ``hot'' PL) have remained unclear~\cite{Orrit-Caldarola_T_measure,metal_luminescence_Link_ASE}, and the interplay between excitation wavelength, resonance position and emission lineshape as well as the differences between the emission and scattering spectra were not well understood (see, e.g.,~\cite{Orrit_QY_metal_PL,metal_luminescence_Link_2011,Baumberg_Narang-ILS,Joel_Yang_ACS_Nano_2012,Joel_Yang_ACS_Photonics_2016,metal_luminescence_Link_SE,metal_luminescence_Link_ASE}). Further, in the majority of cases, the emission was attributed to two-photon interband transitions (usually referred to as two-photon PL, 2PPL, e.g.~\cite{Mooradian_metal_luminescence,Boyd-Shen-luminescence}), while other studies discussed emission occurring within the conduction band (see, e.g,~\cite{Beversluis_PRB_2003,intraband_PL_Xiamen,Baumberg_SERS_T_measure,Baumberg_Narang-ILS}). It was also not clear whether the emission should be associated with the electron or lattice temperatures (e.g.,~\cite{metal_luminescence_Link_ASE}), and there were different claims regarding the scaling of the emission with the electric field amplitude, the nanostructure size, what is the value of the quantum yield, etc.

While some of these arguments recurred also in the context of PL following illumination by ultrafast light pulses (to be denoted below as transient PL, tPL), see, e.g.,~\cite{Boyd-Shen-luminescence,Beversluis_PRB_2003,Bouhelier_PRL_2005,Quidant_modes,Lupton_transient_metal_PL,Hecht_PRL-2019,Suemoto_PRB_2019,Abajo_DTU_PL}, this scenario raised additional specific arguments. Specifically, there is a range of different claims about the electric field dependence of the emission, associated with the {\em integer} number of photons absorbed at the stage preceding the emission. Indeed, there have been reports of linear scaling (see~\cite{Feldman_QY_metal_PL,
Beversluis_PRB_2003,intraband_PL_Xiamen}
), two-photon absorption (2PA)~\cite{Okamoto_2PA_PL,Quidant_modes} (or, as said, 2PPL), three-photon absorption~\cite{Fourkas_NL_PL}, multiple power-laws (2 to 4 photon absorption~\cite{Hecht-Pohl-Science_antenae}, 3 to 5 photon absorption~\cite{Knittel_Leitenstorfer_Brida_PRB_2017}, 4 to 6 photon absorption~\cite{Giessen_SHG_PL}) and even up to 8 photon absorption~\cite{Demichel_PL_2016}. In contrast, some studies explained the emission using a thermal model~\cite{metal_luminescence_Cahill_PNAS,Lupton_transient_metal_PL,Lupton_transient_metal_PL_2017,Bouhelier_JOSAB_2021} such that it is thought of as thermal emission from an object with an optically induced transient electron temperature~\cite{metal_luminescence_Cahill_PNAS,Greffet_PRX_2018,Greffet_Adv_Opt_Mat_2019}. In particular, in~\cite{metal_luminescence_Cahill_PNAS} the emission was interpreted via the ``Two Temperature Model'' (TTM). Such models give rise to {\em non-integer} power-law scaling, see e.g.,~\cite{Lupton_transient_metal_PL,Lupton_transient_metal_PL_2017,Karki_NL_PL}.

The thermal models for the emission came along with the suggestion to spectrally resolve the emission~\cite{metal_luminescence_Cahill_PNAS,Lupton_transient_metal_PL,Lupton_transient_metal_PL_2017,Knittel_Leitenstorfer_Brida_PRB_2017,Abajo_DTU_PL}. Indeed, most earlier experiments measured the total number of emitted photons at all frequencies collected over many pulses. However, the frequency-by-frequency account of the emission in~\cite{metal_luminescence_Cahill_PNAS} was used to show that the emission cannot originate just from 2PA. In~\cite{metal_luminescence_Cahill_PNAS,Lupton_transient_metal_PL,Lupton_transient_metal_PL_2017}, a power-law dependence of the emission on the illuminating field strength was observed experimentally and explained assuming that the emission is purely thermal. In particular, the power-law exponent of the photon emission rate with respect to the incident intensity (or equivalently, the irradiance, i.e., the incident field strength squared) at a particular emission frequency $\omega$ was shown to grow linearly with the emission frequency over a very wide spectral window and for a variety of geometries (random dense films of both gold and silver nanoparticles (NPs) and later for sparse gold NP random arrays~\cite{Lupton_transient_metal_PL_2017}). Similar results were later obtained in~\cite{metal_luminescence_Link_ASE,Bouhelier_JOSAB_2021} for a gold film and Ga spheres~\cite{Ga_spheres}. The model of~\cite{Lupton_transient_metal_PL,Lupton_transient_metal_PL_2017} also explained the growing importance of the blue (i.e., short wavelength) side of the emission upon increase of the incident illumination intensity $I_\text{inc}$. These studies not only further supported the claim that the emission does not originate from 2PA, but they also served to support the claims that the emission originates from intraband recombination events~\cite{Lupton_transient_metal_PL_2017}. Interestingly, however, in~\cite{Lupton_transient_metal_PL_2017} a low intensity measurement revealed a different spectral dependence of the emission on the electric field, namely, a step-like scaling with the incident intensity for the Stokes Emission (SE) and its square for the anti-Stokes Emission (aSE).

These rather different views of the physical nature of metal PL could be associated with three main reasons. First, the experiments were performed for a wide range of structures and illumination conditions. Second, only very few studies resolved the dynamics of the PL spectrally (e.g., as in~\cite{metal_luminescence_Cahill_PNAS,Lupton_transient_metal_PL,Lupton_transient_metal_PL_2017}) or temporally (e.g., as in~\cite{koreans_time_resolved_PL_2002,Feldman_QY_metal_PL,Dantus_2015,Ono_2018,Ono,Suemoto_PRB_2019,Koyama-2021}) and even fewer resolved both; this prevented obtaining a deep understanding of the PL. Third, there was no comprehensive theoretical basis to explain the experimental data. Indeed, although the dynamics of electrons following an ultrashort pulse were well-understood theoretically and experimentally soon after the emergence of femtosecond lasers (e.g.,~\cite{non_eq_model_Lagendijk,delFatti_nonequilib_2000,vallee_nonequilib_2003,Italians_hot_es,GdA_hot_es}), this understanding was not employed to resolve the above disagreements. Specifically, the first attempt to explain the transient PL from metals using a model that relied on the detailed electron dynamics was made in~\cite{Nordlander_transient_e_dynamics}; it relied on the well-established Boltzmann model for the electron dynamics (\cite{delFatti_nonequilib_2000,Italians_hot_es,GdA_hot_es}). The employed approach accounted for the discrete nature of the electron levels (as suitable for 1--2 nm particles~\cite{Khurgin-Levy-ACS-Photonics-2020}). However, it treated the excitation crudely in the sense that its starting point was a maximal deviation from equilibrium due to a pulse with fixed total energy and infinitesimal duration\footnote{However, the photon absorption rate seems to have been vastly overestimated in some of the calculations in~\cite{Nordlander_transient_e_dynamics}. }, rather than a proper modelling of the excitation stage following a pulse with a finite duration. Moreover, while this work computed the $e-e$ transition matrix elements with unusually high accuracy, it treated $e-ph$ interactions phenomenologically, by using the relaxation time approximation, so the total energy of the electron system (hence the intensity of the emission) might not be captured accurately in that work~\cite{Dubi-Sivan,Dubi-Sivan-Faraday}. Progress was made in~\cite{Abajo_DTU_PL} by Echarri {\em et al.} who used a similar model which also accounted for the excitation pulse profile to identify the transition of the aSE from 2PA-based emission to thermal emission (modelled separately via a single temperature model).

A somewhat more accurate model was employed in~\cite{Ono} which used state-of-the-art modelling for the $e-e$ and $e-ph$ interactions as well as photon absorption. The PL was, however, computed approximately by assuming that the electron distribution is thermal with an equivalent electron temperature\footnote{Probably as in~\cite{Italians_hot_es}.} and using the photonic density of states of free space; the latter assumption is suitable for metal nanostructures illuminated away from any resonance, structures having shallow resonances (which indeed includes the Ag film studied in that work) or when the (localized) plasmon resonance (PR) is in the aSE range, but is less suitable for single particles which have a more pronounced plasmon resonance (a scenario abundant in most other experimental work). 
Most recently, Riffe and Wilson's similarly detailed rigorous analysis of the ultrafast dynamics of electrons following a short pulse~\cite{Riffe-Wilson-2022} was applied to the reconstruction of the emission measurements in~\cite{metal_luminescence_Cahill_PNAS}. In that respect, these papers did not address directly the disagreements described above.

Accordingly, the existing work on transient PL from metals does not actually explain if the emission is thermal or non-thermal, and cannot unequivocally determine the scaling of the emission with the electric field. In~\cite{Koyama-2021}, an attempt to separate the non-thermal and thermal component was made. However, this work also treated the excitation stage crudely, and used rather phenomenological expressions for the electron dynamics. 

Below, we rely on recent progress in the understanding of CW PL, and an application of a complete description of the electron non-equilibrium dynamics to answer the open questions associated with the emission statistics and electric-field dependence, demonstrate the findings in new measurements, and use the model to interpret earlier work.

\subsection{Recent progress on CW PL}
Recently, following the progress made in the calculation of the steady-state electron non-equilibrium in metals under CW illumination~\cite{Dubi-Sivan,Dubi-Sivan-Faraday} and its experimental confirmation~\cite{Shalaev_Reddy_Science_2020,Dubi-Sivan-MJs,Gabelli}, Sivan \& Dubi studied the corresponding light emission (PL) from the metal~\cite{Sivan-Dubi-PL_I}; as done frequently~\cite{metal_luminescence_Cahill_PNAS}, this work circumvented the arguments on the exact nature of the emission (i.e., being either a radiative recombination or electronic Raman process) by adopting the perturbative approach and by not specifying the exact operator inducing the emission and its associated matrix elements. Sivan \& Dubi showed that the emission consists of 2 components. The first is {\em thermal} (or, black-body, BB) emission, given by Planck’s Law, which is proportional to the average energy, namely,
\begin{equation}\label{eq:BB}
\sim \langle \e_\text{BB}(\omega,T_\text{e}) \rangle = \frac{\hbar \omega}{e^{\frac{\hbar \omega}{k_\text{B} T_\text{e}}} - 1},
\end{equation}
and which is enhanced by the Purcell effect (via the multiplication by the local density of photonic states, LDOPS, $\rho_\text{phot}(\omega)$); here, $\omega$ is the emission frequency, $k_\text{B}$ is the Boltzmann constant and $T_\text{e}$ is the steady-state electron temperature~\cite{Dubi-Sivan,Dubi-Sivan-Faraday}. This term originates from the thermal part of the electron distribution~\cite{Dubi-Sivan,Dubi-Sivan-Faraday}, regardless of the value of the phonon (lattice) temperature. The second component is {\em non-thermal} emission, which, using the approximate analytic solution derived in~\cite{Dubi-Sivan-Faraday}, was found to consist of a series of Planck-{\em like} terms, namely,
\begin{equation}\label{eq:NTE}
\sim \rho_\text{phot}(\omega) \left[A(\omega;\omega_\text{L},T_\text{e}) \delta_E + B(\omega;\omega_\text{L},T_\text{e}) \delta_E^2 + \cdots \right],
\end{equation}
where
\begin{equation}\label{eq:AB}
A(\omega;\omega_\text{L},T_\text{e}) = 2 \frac{\hbar \left(\omega - \omega_\text{L}\right)}{e^{\frac{\hbar \left(\omega - \omega_\text{L}\right)}{k_\text{B} T_\text{e}}} - 1}, \quad
B(\omega;\omega_\text{L},T_\text{e}) \sim - 2 A(\omega;\omega_\text{L},T_\text{e}) + \frac{1}{2}A(\omega;2\omega_\text{L},T_\text{e}).
\end{equation}
Here, $\omega_\text{L}$ is the frequency of the incoming pump photons and $\delta_E = |\hat{E}(\omega_\text{L})/E_\text{sat}|^2$ is the ratio of the local field and the saturation field (see definition in~\cite{Dubi-Sivan-Faraday,Sivan-Dubi-PL_I}) whose value is $\sim 5$ GV/m for noble metals; $\delta_E$ thus represents the number of non-thermal electrons, those which are not included in the thermal part of the distribution. The shifts by $\omega_\text{L}$ are the signature of the non-thermal (yet, thermal-{\em like}) nature of the emission originating from one photon absorption events, two-photon absorption ({\em sequential}, i.e., uncorrelated) events, etc.

For CW, which naturally involves no more than modestly high intensities and electron temperatures, the $A$ term (first term in Eq.~(\ref{eq:NTE})) is overwhelmingly dominant over all other terms for most experimentally accessible scenarios (i.e., for realistic temperatures and for frequencies not much higher than $\omega_\text{L}$). This makes CW PL a dominantly non-thermal emission effect (i.e., ``hot'' PL). It also has a markedly different spectral signature compared to the thermal emission. Specifically, considering only the electronic contribution to the emission~(see Eq.~(\ref{eq:I_e}) below, obtained by ignoring the LDOPS, or simply normalizing by it), the non-thermal emission alternates between rather flat spectral ranges and those where the emission decays exponentially\footnote{As discussed in~\cite{Sivan-Dubi-PL_I}, this dependence is similar to the step structure seen earlier in the context of ultrafast electron dynamics in, e.g.,~\cite{non_eq_model_Rethfeld,Italians_hot_es}. }; this contrasts with the BB emission which decays exponentially with growing frequency from a maximal value in the deep infrared region, see Fig.~\ref{fig:schematic}.

The analysis in~\cite{Sivan-Dubi-PL_I} enabled the resolution of several long-standing disagreements, e.g., those associated with the statistical nature of the emission, its spectral features and electric-field dependence, the connection to the electron and phonon temperatures etc.

\subsection{Paper outline}
In this work, motivated by the analysis of CW metal PL~\cite{Sivan-Dubi-PL_I}, free of simplifying assumptions such as rapid or effective thermalization~\cite{metal_luminescence_Cahill_PNAS,Lupton_transient_metal_PL}, or neglecting the excitation stage~\cite{Nordlander_transient_e_dynamics,Ono} and free of spectral and intensity limitations of an experiment~\cite{Lupton_transient_metal_PL,Lupton_transient_metal_PL_2017}, we employ a theoretical framework that is capable of answering the various open questions associated with the nature of transient PL from metals. We begin by providing a heuristic explanation of transient PL, see Section~\ref{sec:heuristic}. Then, in Section~\ref{sec:theory}, we describe the theoretical model (Section~\ref{subsec:model}) and calculate numerically the metal PL by solving the Boltzmann-equation for the transient electron dynamics following illumination by a short pulse of a wide range of intensities (Section~\ref{subsub:PL_details}); we then use the extension of the Fermi-golden rule expression for the light emission (derived in~\cite{Sivan-Dubi-PL_I}) to calculate the PL and identify features that can be associated with either non-thermal or thermal effects, thus, enabling the identification of these contributions to the total transient PL.

The novelty of our approach has two components. First, unlike most previous studies in the context of metal PL following illumination by an ultrashort pulse~\cite{metal_luminescence_Cahill_PNAS,Nordlander_transient_e_dynamics,Ono}, we treat the excitation stage rigorously rather than phenomenologically. Second, inspired by the idea of spectrally resolving the emission introduced in~\cite{metal_luminescence_Cahill_PNAS,Lupton_transient_metal_PL,Lupton_transient_metal_PL_2017} and the analysis of the CW emission (which exposed the different contributions to the emission spectrum), we study the evolution of the emission in time. By correlating the distribution dynamics to the emission dynamics and monitoring the evolution of the spectral structure of the emission, we draw conclusions regarding the relative importance of the thermal and non-thermal components of the emission.

In Section~\ref{subsec:E-field-scaling}, we then focus on the scaling of the emission with the electric field. We show that at low intensities, the emission is characterized by a power-law with a step-like exponent, and that this is a signature of PL based on non-thermal electrons. In contrast, at high intensities the power-law attains a smooth linear dependence on the emission frequency. As shown already in~\cite{Lupton_transient_metal_PL,Lupton_transient_metal_PL_2017} for the NP clusters and the dense NP films illuminated by relatively high intensities, this behaviour corresponds to thermal emission. Finally, in Section~\ref{subsec:heating}, we show that cumulative heating (from repeated illumination of one (or more) particles) can explain the decreasing slope of the emission intensity with respect to the illumination power (as reported, e.g., in~\cite{metal_luminescence_Cahill_PNAS,Lupton_transient_metal_PL}).

In Section~\ref{sec:experiments}, we describe new experimental data of tPL from Au rods that demonstrate some of the new aspects of the non-thermal emission stage revealed by our analysis, and discuss the signature of these very aspects in earlier work. In Section~\ref{sec:summary} we provide a summary of our results and an outlook.

\section{Heuristic analysis}\label{sec:heuristic}
Unlike the case of PL under CW illumination~\cite{Sivan-Dubi-PL_I}, an analytic expression for the transient PL is not available. Nevertheless, based on the CW solution (Eqs.~(\ref{eq:BB})-(\ref{eq:AB})), assuming that the electron distribution can be loosely separated into a thermal and non-thermal part (as done frequently in this context~\cite{Stoll_review}), and adopting an ``adiabatic'' point of view (i.e., treating the electron distributions at each time snapshot as if it was a stationary distribution), we expect the emission to be roughly of the following form\footnote{As seen from Eq.~(\ref{eq:AB}), the form adopted here is somewhat simplified and, hence, should not be taken as more than a qualitative description.}
\begin{eqnarray} \label{eq:Gamma_guess}
\Gamma^\text{em}_\text{tot}(\omega) &\sim& \rho_{phot}(\omega) \Big[\langle \e_\text{BB}(\omega;T_\text{e}(t))\rangle + 2 \langle\e_\text{BB}(\omega - \omega_{\text{L},0};T_\text{e}(t))\rangle \delta_E(t) \nonumber \\
&+& \langle \e_\text{BB}(\omega - 2\omega_{\text{L},0};T_\text{e}(t))\rangle \delta_E^2(t) + \cdots\Big].
\end{eqnarray}
Here, $\langle \e_\text{BB}(\omega,T_\text{e}(t)) \rangle$ is, as in Eq.~(\ref{eq:BB}), the Planck (BB) term, i.e., it represents the thermal emission, now at a time-varying electron temperature\footnote{We emphasize that the phonon temperature plays no direct role in the emission, i.e., the thermal emission from the electron-hole recombination is characterized by an electron temperature that may be radically different from the phonon temperature. }; the next terms represent {\em non-thermal} (``hot'') emission, even though they involve frequency-shifted Planck (i.e., thermal-{\em like}) terms; here, $\omega_{\text{L},0}$ is the central frequency of the illuminating laser pulse and $\delta_E = |E(t;\omega_{\text{L},0})/E_\text{sat}|^2$ is the ratio of the (now time-varying) envelope of the local field\footnote{I.e., $\mathscr{E} = E(t;\omega_{\text{L},0}) e^{- i \omega_{\text{L},0} t} + c.c.$.} and the saturation field~\cite{Sivan-Dubi-PL_I}. As in the CW case, the first non-thermal term is proportional to the square of the local electric field, and the emitted frequency is shifted by one photon-energy quantum. This term matches the expression adopted for the electron-hole occupation number in~\cite[Eq.~(2)]{metal_luminescence_Cahill_PNAS}\footnote{Note that this term was interpreted implicitly in~\cite{metal_luminescence_Cahill_PNAS} as having a {\em thermal} origin; our analysis shows that this is (only) thermal-{\em like}. }. The third term is due to two (uncorrelated events of) photon absorption, etc.

Different from the CW case, here, $T_\text{e}$ represents a time-varying effective (/equivalent~\cite{Italians_hot_es}) electron temperature; there are various ways to determine it\footnote{E.g., it can be determined via the total energy of the electron subsystem~\cite{Italians_hot_es,Un-Sarkar-Sivan-LEDD-II}, via energy balance~\cite{GdA_hot_es}, coarse-graining the Boltzmann equation~\cite{Dubi-Sivan-Faraday}, a TTM~\cite{metal_luminescence_Cahill_PNAS} etc.}, which nevertheless all give qualitatively similar behaviour as well as a value which, upon completion of the thermalization of the electron subsystem, emerges to be the actual electron temperature; thus, {\em for the purpose of the current heuristic argument}, these approaches are equivalent. In general, due to the various electron collision mechanisms, the temperature rise is a complicated nonlocal and nonlinear function of the local field. In some simple cases, solving a single temperature equation may suffice, see~\cite{thermo-plasmonics-review}. In these cases, the electron temperature increases above the environment temperature $T_\text{env}$ with the local electric field simply as
\begin{eqnarray} \label{eq:Te_guess}
T_\text{e} &=& T_\text{env} + \Delta T_\text{e}(|E(t;\omega_{\text{L},0})|^2).
\end{eqnarray}

Several insights can already be obtained from Eqs.~(\ref{eq:Gamma_guess})-(\ref{eq:Te_guess}) and are described schematically in Fig.~\ref{fig:schematic}. First, the non-thermal contribution to the emission clearly decays rapidly in time. The heuristic form~(\ref{eq:Gamma_guess}) implies this would occur on the timescale of the local electric field; however, clearly, the actual decay rate would be determined by the thermalization rate (i.e., it would be dominated by $e-e$ collisions). In contrast, the BB term persists for longer times, determined in part by $e-ph$ coupling but mostly by the much slower heat transfer to the environment. Thus, one can expect the early stages of the emission to be non-thermal, but the later (post-thermalization) stages to be of a thermal nature.

Second, following the analysis of the CW case~\cite{Sivan-Dubi-PL_I}, we expect that up to moderately high illumination intensities (for which there is negligible light-induced heating), the non-thermal emission would be far stronger than the thermal emission; indeed, the latter is generally weak and peaks at the infrared regime. The thermal emission, however, will be gradually stronger with growing illumination intensity, because the (electron and overall) heating would be more significant and thus make the thermal component stronger and blue-shift it into the visible frequency range. In a standard measurement, the emission is time-integrated, so the statistical nature of the emission will depend on the relative importance of these two contributions.

Lastly, it becomes obvious (e.g., following~\cite[Fig.~3(b)]{Sivan-Dubi-PL_I}) that up to moderately high illumination intensities, the dependence of the PL on the electric field is simply polynomial, corresponding to the $m$-photon absorption terms. In particular, the 2PA term is expected to dominate the aSE above some frequency; at $T_\text{e} \to 0$, this frequency is simply $\omega_{\text{L},0}$, and it grows with $T_\text{e}$. Going further, the 3PA term dominates above $2 \omega_{\text{L},0}$, and so forth. Correspondingly, the transitions between the spectral regions dominated by $m$-photon absorption, and ($m+1$)-photon absorption, smear out for growing $T_\text{e}$. However, upon significant heating, the polynomial description fails due to the (exponential) dependence of the coefficients $\langle \e(\omega - m \omega_{\text{L},0};T_\text{e})\rangle$ on the electric field (see Eq.~(\ref{eq:Te_guess})). In both cases, the nonlinearity of the emission is electronic in nature, and associated primarily with the transient effect caused by each individual absorbed pulse; the increase of the phonon and environment temperatures is much smaller, and accumulates only on the timescale of many pulses, see discussion below in Section~\ref{subsec:heating}. Therefore, the temperature rise of the phonons and environment has at most a modest quantitative effect on the emission.

Below, we show numerical simulations that support these heuristic expectations. They, however, also show that the heuristic solution~(\ref{eq:Gamma_guess}) captures the dynamics only qualitatively, so~(\ref{eq:Gamma_guess}) will serve only for the purpose of distinguishing the crossover from non-thermal (absorption-induced) emission to thermal emission.

\section{Theory of transient PL}\label{sec:theory}
\subsection{Rigorous model}\label{subsec:model}

Determining the transient PL requires first knowing the transient electron distribution dynamics, $f(\e,t)$ (with $\e$ being the electron energy). We determine $f$ by solving the time-dependent Boltzmann equation assuming that electron states in the metal are characterized by a continuous energy variable. In Refs.~\cite{Kreibig_PWA_epsilon_model,Khurgin-Levy-ACS-Photonics-2020}, it was shown that for NP sizes as small as $\approx 2$~nm the results of this approach are in excellent agreement with the more accurate discretized momentum-space models.

The electron interactions are accounted for using a standard model, not much different from the one used in the original studies of the problem (e.g.,~\cite{delFatti_nonequilib_2000,Italians_hot_es,GdA_hot_es,Stoll_review}), incorporating rigorously electron collision mechanisms and especially the photon absorption events, see Appendix~\ref{app:model}\footnote{Since the electron dynamics were studied systematically in many previous articles~\cite{delFatti_nonequilib_2000,Italians_hot_es,Stoll_review,Ono,Riffe-Wilson-2022}, we do not show them here explicitly. }. In particular, we account for intraband absorption events as well as interband transitions by accounting for the empirical value used for the imaginary part of the permittivity in Poynting's Theorem\footnote{Thus, we do not account for two simultaneous photon absorption events from the $d$ (‘valence’) band to the conduction band. As shown below, we are not convinced that earlier claims for the dominance of these transitions are of general validity. }. This rigorous treatment of the absorption enables us to study the early stages of the PL dynamics, and hence, to study both the statistical aspects of the emission as well as its dependence on the electric field. This is the main distinction of our work from earlier theoretical studies of the transient emission from metals~\cite{Nordlander_transient_e_dynamics,Ono}.

As shown in~\cite{Lupton_transient_metal_PL,Lupton_transient_metal_PL_2017,Bouhelier_JOSAB_2021}, the emission has similar characteristics for different particle geometries; thus, except for its effect on the LDOPS (see below), the nanostructure geometry in our model manifests itself only in the connection of the local field to the incident field. In that respect, we consider only the averaged electric field inside the metal, and rely on the strong electron diffusion~\cite{ICFO_Sivan_metal_diffusion} to justify the assumption that not only is the temperature uniform in the nanostructure, but also the electron distribution itself\footnote{To the best of our knowledge, there is to date no formulation that treats field and electron temperature non-uniformities in illuminated metal nanoparticles in a rigorous manner that can enable going beyond the assumptions of uniformity and field averaging. First steps towards this goal were performed in~\cite{NESSE,Moloney_PRB_2021}. The assumptions of uniformity certainly hold beyond the first few 10s of femtoseconds or so. The emission occurring before this stage is a very small fraction of the total emission, hence, whatever inaccuracies are associated with these assumptions are expected to be small, and not to affect any of the results in the current manuscript.}.

We verified that even for the most intense illumination level used in this work, the emission is much weaker compared to all the effects accounted for above when calculating the distribution $f(\e,t)$. Thus, the emission can be determined from the distribution via a perturbative calculation (as done previously for CW PL in~\cite{Sivan-Dubi-PL_I}, using time-dependent quantum mechanical perturbation theory (Fermi golden rule)). In particular, in analogy to the steady-state case, the transient PL (from a point in the metal NP) is given by~\cite{Sivan-Dubi-PL_I}
\begin{eqnarray}\label{eq:tPL}
\Gamma^\text{em}(\vec{r},\omega,t) &=& \frac{\pi \omega V_\text{NP}^2}{\epsilon_0} \rho_\text{phot}(\vec{r},\omega) I_\text{e}(\omega,t),
\end{eqnarray}
where $V_\text{NP}$ is the NP volume, $\epsilon_0$ is the vacuum permittivity, and $\rho_\text{phot}$ is the LDOPS which is set, for simplicity, to have a Lorentzian-like spectrum centered at $\omega_\text{PR}$ and width of $\gamma_\text{PR} = \omega_\text{PR}/10$. In that sense, we discuss only single-mode emission; it is natural to refer to this mode as the dipolar one, as it is the observable by far-field measurements. This also allows us to suppress the intricate dependence of the LDOPS on the spatial coordinate. Finally, the electronic contribution to the emission is given by
\begin{equation}\label{eq:I_e}
I_\text{e}(\omega,t) = \int_0^{\e_{max}} |\vec{\mu}(\e,\e + \hbar \omega)|^2 \rho_\text{J}(\e,\e + \hbar \omega,t) d\e,
\end{equation}
where $\e_{max}$ represents the top of the conduction band, $\vec{\mu}$ is the dipole moment transition matrix element and $\rho_\text{J}$ is the population-weighted joint density of pair states, given by
\begin{eqnarray}\label{eq:rho_J}
\rho_\text{J}(\e_\text{f},\e_\text{i},t;\omega_\text{L,0}) = \left\{f(\e_\text{i},t;|E(\omega_\text{L,0})|^2) \rho_\text{e}(\e_\text{i})\right\} \left\{\left[1 - f(\e_\text{f},t;|E(\omega_\text{L,0})|^2) \right] \rho_\text{e}(\e_\text{f})\right\}.
\end{eqnarray}
Here, $\e_\text{i} = \e_\text{f} + \hbar \omega$ are the initial and final electron energies involved in the emission of a $\omega$ photon.

We note that the final state $\e_\text{f}$ can, in principle, be either in the conduction band, or in the $d$ (‘valence’) bands (see, e.g.,~\cite{Rosei_Ag_diel_function,Stoll_review} for more details on metal band structure); the corresponding matrix elements for these two processes are comparable\footnote{Indeed, converting the value reported in~\cite{Biancalana_NJP_2012} for the expectation value of the momentum operator to the dipole matrix element and ignoring the degeneracy gives $\sim 1.8 \cdot 10^{-29}$ C$\cdot$m (for an interband transition occurring near the X point of the band structure of the metal), whereas the value obtained in~\cite{Khurgin-Levy-ACS-Photonics-2020} for an intraband transition of energy $\hbar \omega = 1.8$ eV to a state away from the Fermi level is $3.81 \cdot 10^{-29}$~C$\cdot$m$\cdot\frac{\text{nm}}{L}$  where $L$ is the particle size.}. In the latter case, the hole occupation is essentially negligible~\cite{Rosei_Ag_diel_function,Stoll_review}, a fact which immediately shows that the contribution of transitions to the lower $d$ electron bands is negligible with respect to the intraband PL. This conclusion is in line with the reports of~\cite{Lupton_transient_metal_PL,Lupton_transient_metal_PL_2017,Bouhelier_JOSAB_2021} that the (t)PL has similar characteristics for different materials (Au and Ag) and justifies focusing on the electron distribution in the conduction band only.

Where necessary, in order to mimic the experimental data~\cite{metal_luminescence_Cahill_PNAS,Lupton_transient_metal_PL,Lupton_transient_metal_PL_2017}, we integrate the emission over time (for a single excitation pulse). In this case, the total emission is given by
\begin{equation}
I_\text{e,tot}(\omega) = \int_0^{1/f_\text{rep}} I_e(\omega,t) dt, \quad \Gamma^\text{em}_\text{tot}(\omega) = \int_0^{1/f_\text{rep}} \Gamma^\text{em}(\omega,t) dt,
\end{equation}
where $f_\text{rep}$ is the pulse repetition rate.

\begin{figure}[h]
\centering{\includegraphics[scale=0.55]{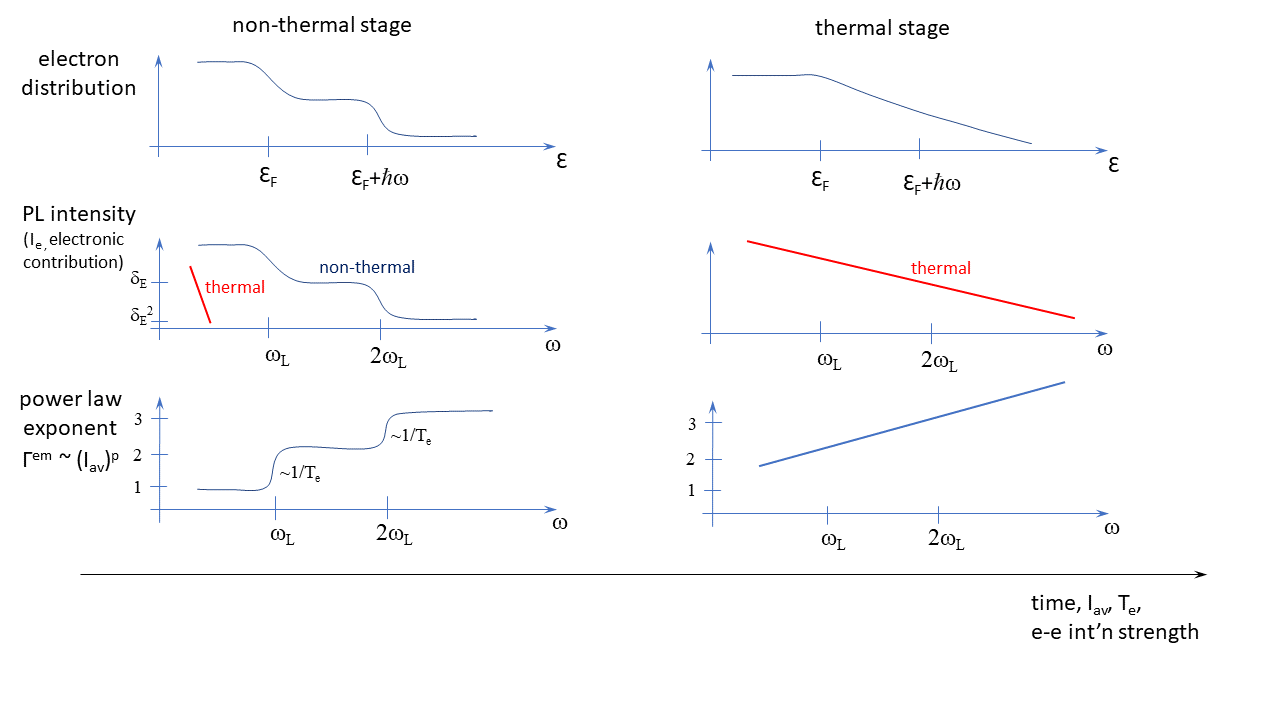}}
\caption{A schematic of electron distribution (top row), the electronic contribution to the (near infrared and visible) emission~(\ref{eq:I_e}) and the power-law exponent (bottom row) for the non-thermal stage (left) and the thermal stage (right) of the emission. The transition between these two cases occurs as the dynamics progresses in time, or for higher (average) illumination intensity ($I_{\text{av}}$), higher electron temperature, or for stronger $e-e$ interactions.} \label{fig:schematic}
\end{figure}

\subsection{Numerical results}\label{subsec:numerics}



Following~\cite{Lupton_transient_metal_PL_2017}, we assume that the Au rod studied in~\cite{Lupton_transient_metal_PL_2017} is illuminated by a $\tau_\text{L} = 85$~fs-long (Gaussian) laser pulse with a central frequency $\omega_{\text{L},0}$ corresponding to a wavelength of $700$~nm, incident intensity given by $I_\text{inc}(t) = I_0 e^{- 4 \ln 2 t^2/ \tau_\text{L}^2}$ and average intensity varying between $I_{\text{av}} = I_0 \tau_\text{L} f_\text{rep} \sim 0.1 - 10$~kW/cm$^2$, with $I_0$ being the peak intensity and $f_\text{rep} = 80$ MHz being the pulse repetition rate (i.e., starting at somewhat lower values and going up to somewhat higher values compared to those used in~\cite{Lupton_transient_metal_PL_2017}). 
Notably, the population of high energy electron states never exceeds a few percent even for the highest illumination intensities used.

\subsubsection{The distribution and emission dynamics}\label{subsub:PL_details}
\begin{figure}[h]
\centering{\includegraphics[width=18cm,height=14cm]{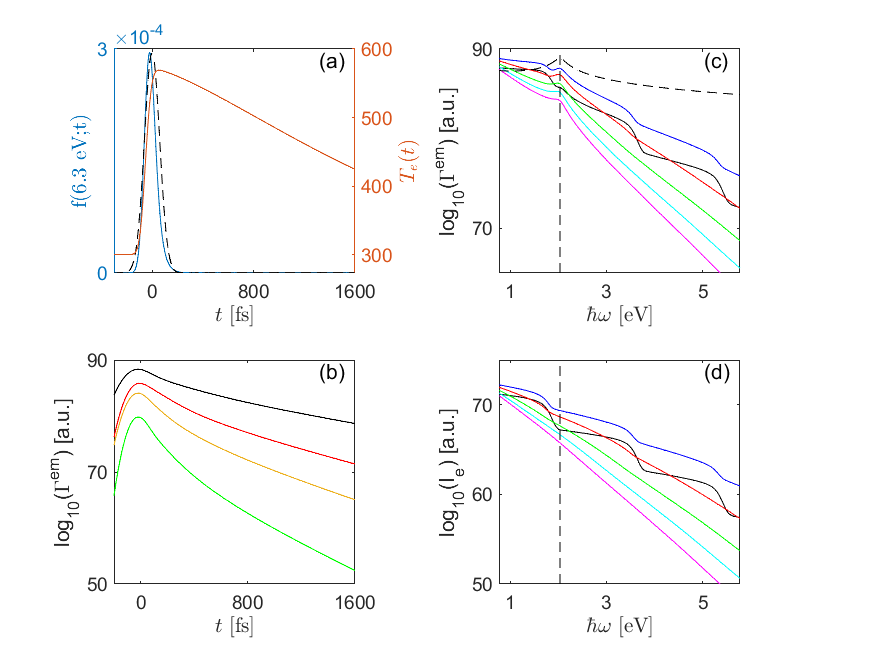}}
\caption{Metal PL following illumination by a 0.5~pJ ($I_{\text{av}} = 0.1$~kW/cm$^2$), 85~fs-long pulse centered at $\hbar \omega_{\text{L},0} = 1.77$~eV. (a) Dynamics of the electric field (dashed black line), electron distribution (at the high energy of $6.3$~eV, blue) and effective electron temperature (orange). 
(b) $\text{log}_{10}[\Gamma^\text{em}(\hbar \omega = 1.05, 1.77, 2.36, 3.28\text{~eV};t)]$ shown by black, red, orange and green solid lines; the emission at the higher frequencies is naturally weaker, but also decays faster due to FLT. (c) PL spectra at different times, $\text{log}_{10}[\Gamma^\text{em}(\omega;t = -100, 0, 100, 200, 300, 400\text{~fs})$], shown in black, blue, red, green, cyan and magenta solid lines. (d) Same for $I_\text{e}(\omega,t)$. The vertical dashed black line in the plots shows the position of the plasmon resonance (PR) at $\hbar \omega_\text{PR} = 1.15 \omega_{\text{L},0} \cong 2$~eV.} \label{fig:emission_0_5pJ}
\end{figure}

We first show the calculated distribution ($f(\e,t)$; Fig.~\ref{fig:emission_0_5pJ}(a)) and PL ($\Gamma^\text{em}(\omega,t)$; Fig.~\ref{fig:emission_0_5pJ}(b)-(c)) for a moderately high energy pulse (0.5~pJ, equivalent to $I_{\text{av}} = 0.1$ kW/cm$^2$); for convenience, cross-sections of the electronic contribution to the emission ($I_\text{e}$, Eq.~(\ref{eq:I_e})) are shown as well (Fig.~\ref{fig:emission_0_5pJ}(d)). 

{\em Dynamics.} We observe that the emission intensity initially grows together with the illuminating pulse intensity and the high-energy non-thermal electron occupation probability (see Fig.~\ref{fig:emission_0_5pJ}(a)-(b)); the emission peaks at a slight delay of a few tens of femtoseconds with respect to the pulse, together with the peak of the high-energy non-thermal electron occupation.

Beyond this stage, the PL intensity decreases with time at a rate that is a non-trivial combination of both $e-e$ and $e-ph$ interactions (see related discussions, e.g., in~\cite{non_eq_model_Lagendijk,Stoll_review,Suemoto_PRB_2019,Wilson_Coh,Ono,Riffe-Wilson-2022}). 
Overall, the decay rate is on the order of a few hundred femtoseconds. In that sense, it is slower compared to the pulse duration\footnote{This discrepancy illustrates the inaccuracy of the heuristic analysis of Section~\ref{sec:heuristic}. } and the decay of the high energy electron occupancy (determined by $e-e$ collisions), but commensurate with the decay rate of the effective electron temperature (determined by $e-ph$ collisions) as deduced from the total energy in the electron subsystem 
(see Fig.~\ref{fig:emission_0_5pJ}(a)). 

A more careful look at these data sets shows that the emission of high-frequency photons decays faster than the emission of lower-frequency photons (Fig.~\ref{fig:emission_0_5pJ}(b)); this observation matches the experimental observation in~\cite[Fig.~1]{Suemoto_PRB_2019}. This is also in line with Fermi's liquid theory (FLT)~\cite{Quantum-Liquid-Coleman}, which predicts that the electron collision rate is faster the further away the energy is from the Fermi energy, i.e., $\tau_{\text{e-e}}^{-1} \sim \left[(\pi k_\text{B} T_\text{e})^2 + (\e - \e_\text{F})^2\right]$. Indeed, as shown already in~\cite{Fann_1992} and later in e.g.,~\cite{Cluzel_PL_2016} in the specific context of tPL, the occupation of higher-energy electrons decays faster than that of lower-energy electrons, thus causing high frequency emission to decay faster than low frequency emission. Being a characteristic of the transition from a non-thermal to a thermal distribution, this observation further supports the interpretation of the peak (i.e. early emission) as non-thermal light emission. Specifically, the lower frequency emission (black line in Fig.~\ref{fig:emission_0_5pJ}(b)) decays on a scale commensurate with the electron temperature decay (orange line in Fig.~\ref{fig:emission_0_5pJ}(a)); this happens due to electron cooling via $e-ph$ interactions. In contrast, the emission at higher frequencies initially decays more rapidly, in correlation to the decay dynamics of the occupation (blue line in Fig.~\ref{fig:emission_0_5pJ}(a)), as predicted by Fermi's liquid theory; the post-thermalization stages of the emission, however, occur on the timescale of the electron temperature decay, as expected.

As discussed in~\cite{Ono,Riffe-Wilson-2022}, the dynamics are affected only in a modest quantitative manner by somewhat different $e-e$ interaction strengths (compare~\cite[Figs.~4(b)-(c)]{Ono}). For the purpose studied here, stronger $e-e$ interactions would make the thermal emission component relatively stronger compared to the non-thermal part (see Fig.~\ref{fig:schematic}). Similarly, there are only modest quantitative changes for longer illumination pulses, amounting to a smearing of the electron distribution and PL dynamics.

However, the dynamics are modified more significantly when the illumination intensity increases. Under such conditions, the thermalization occurs more rapidly~\cite{Fann_1992,Stoll_review,Suemoto_PRB_2019,Ono}. As a result, the contribution to the emission from the late (thermal) stages of the dynamics becomes relatively more important for strong illumination compared to moderate illumination. Because of all this, at the later stages of the emission and for higher excitation intensities, the emission should be thought of as dominantly thermal light, in agreement with the interpretation in~\cite{metal_luminescence_Cahill_PNAS}\footnote{Obtained by studying emission from longer pulses than ours (0.45 and 2 ps).} and~\cite{Lupton_transient_metal_PL,Lupton_transient_metal_PL_2017,Bouhelier_JOSAB_2021}. Simply put, the reason for the stronger thermal characteristic at higher intensities is that under these conditions, $T_\text{e}$ is higher, so the thermal component is correspondingly more significant.

\begin{figure}[h]
\centering{\includegraphics[width=18cm,height=14cm]{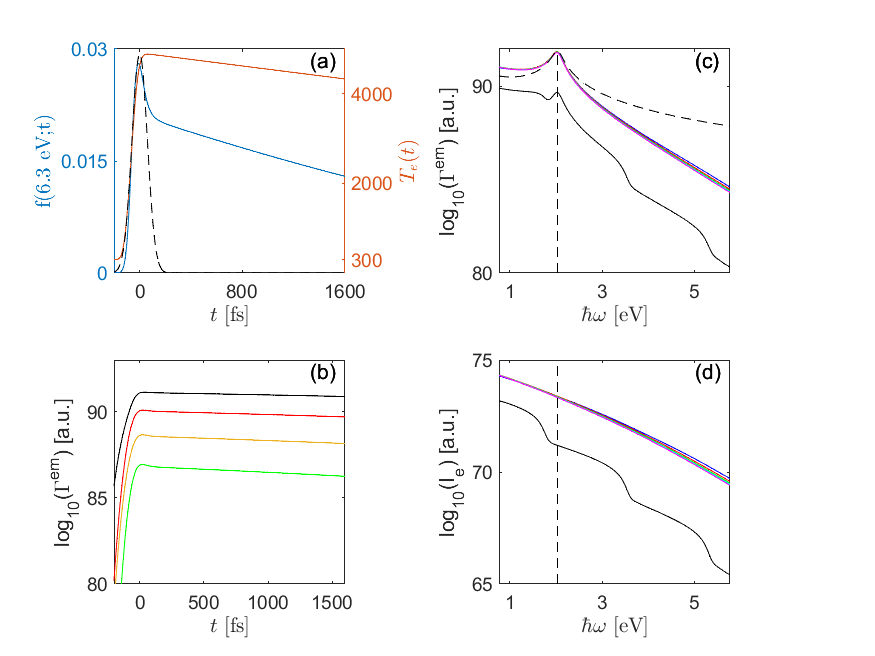}}
\caption{Same as in Fig.~\ref{fig:emission_0_5pJ} for an incident pulse energy of 50~pJ ($I_{\text{av}} = 10$~kW/cm$^2$). The almost immediate emergence of thermal characteristics in the emission is apparent. } \label{fig:emission_50pJ}
\end{figure}

One should also note that the thermal component of the emission persists until the arrival of the next pulse in the pulse train, i.e., far longer than the sub-picosecond extent of the non-thermal emission. This is particularly significant for the aSE (e.g., the green curve in Fig.~\ref{fig:emission_0_5pJ}(b)) and for high intensities (Fig.~\ref{fig:emission_50pJ}(b)), for which the peak of the emission (the non-thermal component) is not much higher than the late emission (i.e., the thermal component); thus, when the emission is integrated in time, the orders-of-magnitude longer extent of the thermal emission may convey a higher relative importance of thermal emission compared to a time-resolved data acquisition. In that respect, the repetition rate may affect the conclusion about the nature of the emission if judged via the time-integrated spectrum.

{\em Emission spectra.} As shown in Fig.~\ref{fig:emission_0_5pJ}(c)-(d) (also compare to Fig.~\ref{fig:schematic}), the spectrum of the emitted light (normalized by the LDOPS) in the initial stages of the dynamics exhibits the step-like structure observed in the CW case~\cite{Sivan-Dubi-PL_I}; in particular, the emission spectrum is determined by the LDOPS, but its blue side is quenched by the electronic contribution, $I_\text{e}$. This structure, however, gets gradually smeared into the monotonically-decreasing spectra characteristic of thermal emission; this is seen also in the time-integrated spectra of Fig.~\ref{fig:SB_power_law_analysis}(a).

An additional aspect of the transient emission, which has been under debate, is the dynamics of the spectral lineshape. Unlike the claim in~\cite{Nordlander_transient_e_dynamics}, we observe in Fig.~\ref{fig:emission_0_5pJ}(c) and Fig.~\ref{fig:emission_50pJ}(c) that the peak of the emission spectrum does not vary in time. Again, this is seen also in the time-integrated spectra of Fig.~\ref{fig:SB_power_law_analysis}(b). However, the red (i.e., low frequency) side of the spectrum decays more slowly than the blue (high frequency) side. This behaviour can, again, be traced to the overall decrease in the occupation of high-energy electron states, resulting in an overall red-shift of the spectra. Notably, for systems with a flat resonance (such as films), this effect manifests itself as a continuous red-shift of the spectrum, see the measurements in~\cite{Suemoto_PRB_2019}. Fig.~\ref{fig:SB_power_law_analysis}(b) also shows that the blue (aSE) side of the emission spectrum broadens with growing illumination intensity, as predicted and observed in~\cite{Lupton_transient_metal_PL,Lupton_transient_metal_PL_2017}. The reason for this is that the aSE is connected to the thermal part of the electron distribution and of the emission (see discussion in~\cite{Baumberg_SERS_T_measure,Cahill_T_measure,Orrit-Caldarola_T_measure,Sivan-Dubi-PL_I}), which naturally grows with the illumination intensity\footnote{This behaviour was referred to as a spectral blue-shift~\cite{Lupton_transient_metal_PL,Lupton_transient_metal_PL_2017,Nordlander_transient_e_dynamics}; indeed, the moments of the emitted spectra blue-shift. }.

\subsubsection{Quantifying the dependence on the electric field}\label{subsec:E-field-scaling}

\begin{figure}[h]
\centering{\includegraphics[width=18cm,height=14cm]{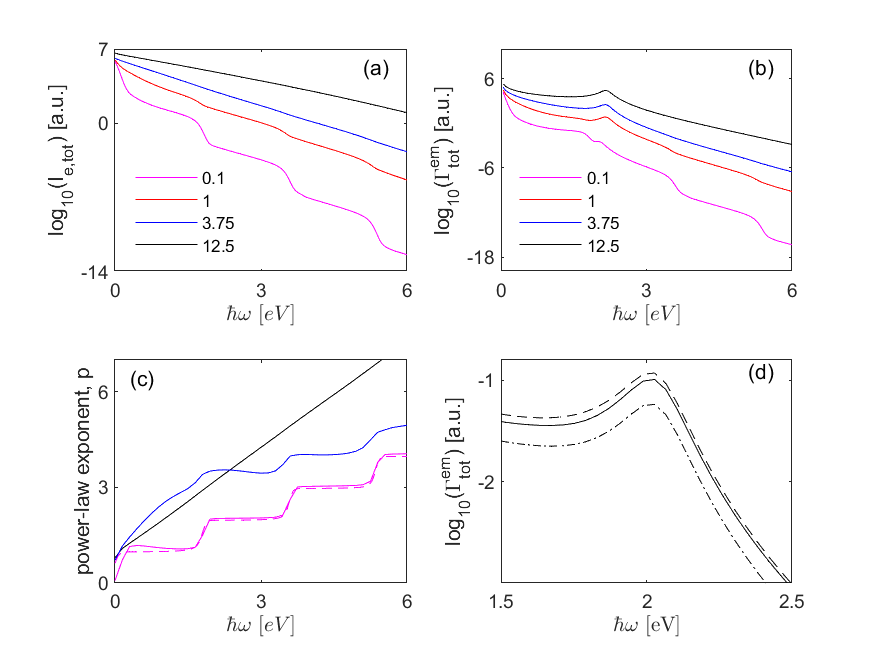}}
\caption{Time-integrated spectra of (a) the electronic contribution ($I_{e,tot}$) and (b) total emission spectra ($\Gamma^{em}_{tot}$) as a function of emission frequency for different illumination intensities in the range spanned by the levels in Figs.~\ref{fig:emission_0_5pJ}-\ref{fig:emission_50pJ}; all other illumination parameters are the same as in those figures. The legend corresponds to the values of the average incident illumination intensity $I_{\text{av}}$ in kW/cm$^2$. (c) Power-law exponent extracted from the rigorous numerical simulations for various excitation intensities (0.8 -- 1.25~pJ in magenta; 12 -- 18.75~pJ in blue; 40 -- 62.5~pJ in black); for the lower pulse energy, we also show the power-law obtained from the electron scattering free analysis described in Appendix~\ref{app:power-law-exponent}. (d) Time-integrated emission spectra for an illuminating pulse energy of 62.5~pJ ($I_{\text{av}} = 12.5$~kW/cm$^2$). The spectra accounting for 300~K and 550~K background temperatures are shown by the solid and dashed lines, respectively, and the dynamics accounting also for the reduced quality factor (due to the increased imaginary part of the permittivity) is shown by the dash-dotted lines. The plot shows only the spectral range near the emission peak. } \label{fig:SB_power_law_analysis}
\end{figure}

The time-integrated emission spectra shown in Fig.~\ref{fig:SB_power_law_analysis}(a)-(b) enable the extraction of the dependence of the tPL on the electric field strength. In particular, following~\cite{Lupton_transient_metal_PL,Lupton_transient_metal_PL_2017} (see Appendix~\ref{app:power-law-exponent}), we linearly fit a double-logarithmic representation of the emission $\Gamma_\text{tot}^\text{em}$ as a function of irradiance $I_{\text{av}}$ for each frequency in the emission spectrum, i.e., we determine the spectrum of the power-law coefficient $p$ in $\Gamma_\text{tot}^\text{em}\sim \left(I_{\text{av}}\right)^p$. For this purpose, typically only a narrow range of irradiances is used such that $p$ constitutes an effective coefficient of nonlinearity {\em local} in both emitted photon energy and illuminating irradiance.

In Fig.~\ref{fig:SB_power_law_analysis}(c) we show the results of the application of the power-law extraction algorithm of~\cite{Lupton_transient_metal_PL,Lupton_transient_metal_PL_2017} on the numerical data shown in Section~\ref{subsub:PL_details}. For only {\em moderately high} illumination intensities, one can see a rather clear step structure. Specifically, we obtain $p = 1$ for the SE and $p = 2$ for the low frequency part of the aSE (i.e., frequencies only modestly above $\omega_{\text{L},0}$); at emission frequencies higher than $2 \hbar \omega_{\text{L},0}$, we observe an additional step of $p = 3$, which has not been reported before. The general staircase-like structure of this spectrum of power-law coefficients is also evident in simulations devoid of electron-electron and electron-phonon interaction, see Appendix~\ref{app:stepedges}. These observations are in line with the predictions of the heuristic model (Section~\ref{sec:heuristic}) and hence, is a clear signature of the dominance of the tPL by non-thermal emission. Remarkably, this step-like behaviour also matches the experimental observations of tPL in~\cite[Fig.~3(b)]{Lupton_transient_metal_PL_2017} for low illumination intensity from a single Au nanorod; for the aSE, a $p = 2$ exponent was also observed in~\cite[Fig.~2B]{metal_luminescence_Cahill_PNAS}\footnote{Our model also matches the observation in~\cite{Feldman_QY_metal_PL} of a $p = 1$ exponent for the integrated spectrum, which is naturally dominated by the SE.}. Notably, this power-law is attained even in the absence of interband transitions, so the popular attribution of the PL nonlinearity to simultaneous two-photon interband transitions, dating back as early as~\cite{Mooradian_metal_luminescence,Boyd-Shen-luminescence}, does not seem to be of  general validity. Instead, one should understand the result as an effective nonlinearity averaged over multiple sequential electron excitation pathways, as pointed out by Knittel {\em et al.}~\cite{Knittel_Leitenstorfer_Brida_ACSNano_2015}.

Spectrally narrow transitions between these regimes of constant nonlinearity are found at frequencies near integer multiples of the excitation laser frequency. The shape of these transitions is a representation of the electron distribution existing before the arrival of the laser pulse and can be used to determine the average lattice temperature. With increasing illumination intensity, the transitions distinctly shift towards lower photon energies, as detailed in Appendix~\ref{app:stepedges}, and the staircase-like pattern in the power-law exponent spectrum gradually evolves into a straight line. The latter behaviour is a clear signature of thermal emission, as it was shown in~\cite{Lupton_transient_metal_PL} that for thermal electron distributions, the emission is given by\footnote{One can naturally also rewrite Eq.~(\ref{eq:linear_p}) in terms of the average intensities.}
\begin{equation}
\log\left[\e_\text{BB}(I_0)) / \e_\text{BB}(I^*_0))\right] \sim p(\hbar \omega) \log[I_0/I^*_0], \label{eq:linear_p}
\end{equation}
with $p(\hbar \omega) = \hbar \omega /(a k_\text{B} T^*)$;this relation can be derived for a narrow range of illumination intensities around a reference illumination intensity $I^*_0$ and reference electron temperature $T^*$ and the approximation that the illumination intensity-dependence and the electron temperature are related via $(T⁄T^*)^a = I_0⁄I^*_0$ as expected for an electron gas. Here, $a$ is an order unity dimensionless number\footnote{This is indeed the high temperature limit of Eq.~(\ref{eq:linear_p}) for $a = 2$. }. Remarkably, this transition happens for low frequencies (SE) at relatively low intensities and for high frequencies (aSE) at higher intensities. This agrees with the above heuristic interpretation (Section~\ref{sec:heuristic}), namely, that the thermal contribution becomes stronger and shifts from the infrared into the visible spectral range as the illumination intensity grows. For a more detailed description of the power-law results, see Appendix~\ref{app:simulatedpowerlawspectra}.

We note, however, that for high emission frequencies, the power-law exponent may undergo some changes once additional physical effects shall be taken into account. The reason for this is that the emission at such frequencies would not only depend on the exact population at low electron energies in the conduction band, but also on the population of the various d ('valence') bands which are not accounted for in the current study. Moreover, the high frequency emission will be limited by the band edge (vacuum level).

\subsubsection{The role of cumulative (steady-state) heating and permittivity changes}\label{subsec:heating}
In all our calculations, we neglected the small transient  changes to the permittivity following each pulse absorption event~\cite{Stoll_review}. However, even though the absorbed heat is mostly transferred to the environment before the arrival of the next pulse, a small fraction of it increases the electron and phonon baseline temperature of the NP (and environment) slightly. Thus, it is clear that the temperature of the system will gradually grow due to accumulation of this residual absorbed energy after a sufficiently large number of pulses~\cite{Y2-eppur-si-riscalda,Un-Sivan-sensitivity}. This effect becomes stronger in case of heating from adjacent illuminated NPs~\cite{Y2-eppur-si-riscalda,Un-Sivan-sensitivity,Baffou-Quidant-Baldi}. Such long-term cumulative heating (of the electron, phonon and environment temperatures) by several hundreds of degrees gives rise to a thermo-optic nonlinearity, namely, it is expected to cause changes to the (steady-state) metal permittivity~\cite{Shalaev_ellipsometry_gold,Shalaev_ellipsometry_silver,PT_Shen_ellipsometry_gold} and, hence, to reduce the local field strength~\cite{Sivan-Chu-high-T-nl-plasmonics,Gurwich-Sivan-CW-nlty-metal_NP,IWU-Sivan-CW-nlty-metal_NP}, such that the steady-state temperature rise will become sublinear with respect to the irradiance. Eventually, the heating will cause surface melting and volume changes, particle reshaping and ultimately sample damage~\cite{japanese_size_reduction,orrit_size_reduction}.

In order to determine this long-term steady-state temperature rise via numerical simulations, one needs to know the details of the nanostructure geometry, the pulse repetition rate, the outer thermal boundary conditions, etc. (see, e.g.,~\cite{Un-Sivan-sensitivity}). Instead, for simplicity, we estimate the cumulative heating in the specific case of~\cite{Lupton_transient_metal_PL_2017} as follows: We assume that the sample reaches the temperature threshold for damage/sintering (estimated to be $T_\text{env} \sim 550$~K~\cite{PT_Shen_ellipsometry_gold}) due to cumulative heating for average illumination intensities of $I_{\text{av}} = 12.5~\text{kW}/\text{cm}^2$ (i.e., slightly exceeding the highest used in~\cite{Lupton_transient_metal_PL_2017}). Fig.~\ref{fig:SB_power_law_analysis}(d) shows that this causes a slight increase of the emission intensity in comparison to the case of $T_\text{env} = 300$~K. Indeed, at the initial stages of the dynamics, the aSE (which is associated with the thermal part of the emission~\cite{Sivan-Dubi-PL_I}) is higher, and as the system thermalizes, the SE intensity also increases slightly (due to the thermalization to a slightly higher temperature).

However, for resonant illumination, this slight increase in emission intensity is overwhelmed by the stronger (thermo-optic) effect - the imaginary part of the metal permittivity increases~\cite{Shalaev_ellipsometry_gold,Shalaev_ellipsometry_silver,PT_Shen_ellipsometry_gold}\footnote{An exception is the regime of wavelengths shorter than $\sim 500$~nm, which is dominated by interband transitions~\cite{Stoll_environment,PT_Shen_ellipsometry_gold}. }, and consequently, the quality factor of the plasmon resonance drops and the local field decreases~\cite{plasmonic-SAX-PRL,Sivan-Chu-high-T-nl-plasmonics,Gurwich-Sivan-CW-nlty-metal_NP,IWU-Sivan-CW-nlty-metal_NP}. As a result, the emission intensity decreases across the entire spectrum (again, see Fig.~\ref{fig:SB_power_law_analysis}(d)). As the illumination is shifted away from resonance, this effect becomes gradually weaker (not shown).


In this respect, the thermo-optic effect should thus be taken into account when explaining the experimental observations of a decreasing power exponent of the aSE in the high-power range in~\cite[Fig.~2B]{metal_luminescence_Cahill_PNAS} and in~\cite{Lupton_transient_metal_PL}\footnote{Note that a similar effect occurs due to the temperature dependence of the electron heat capacity~\cite{Ashcroft-Mermin,Lupton_transient_metal_PL}; however, such an effect is associated with the ultrafast electron dynamics, rather than the steady-state heating that is discussed in the current section. }. 
Our explanation 
refines the suggestion in~\cite{metal_luminescence_Cahill_PNAS} that $\sigma_\text{abs}$ grows for high electron temperatures. The latter claim was shown in~\cite{Sivan-Chu-high-T-nl-plasmonics,Gurwich-Sivan-CW-nlty-metal_NP,IWU-Sivan-CW-nlty-metal_NP} to hold only for off-resonance illumination.




\section{Experiments}\label{sec:experiments}



The linear power-law spectrum $p \sim \hbar \omega$ discussed in Section~\ref{subsec:E-field-scaling} has been observed by several groups for metal NPs excited by ultrashort laser pulses~\cite{Lupton_transient_metal_PL,intraband_PL_Xiamen,Ga_spheres,metal_luminescence_Link_ASE
} and can be understood reasonably well by the analysis of a thermalized electron gas~\cite{Lupton_transient_metal_PL} as discussed above. On the other hand, the transition between the low-irradiance, staircase-like and the high-irradiance, linear power-law spectrum has so far only been reported once~\cite{Lupton_transient_metal_PL_2017} and the physical nature of the underlying nonlinearity was not fully understood.

The model presented so far attributes the low-irradiance nonlinearity to the effective number of photon excitations necessary to excite the electronic states contributing to the joint density of states $\rho_\text{J}$~(Eq.~(\ref{eq:rho_J})), even when the transitions happen sequentially within the (same) conduction band (see Appendix~\ref{app:stepedges}). Two critical predictions can be derived that so far lack experimental confirmation: first, moderate-irradiance power-law spectra acquire a staircase-like structure with integer values for the power-law coefficient, even beyond the value of $p = 2$ demonstrated in~\cite{Lupton_transient_metal_PL_2017}. Second, transitions between the spectral regions of integer power-law coefficients $m$ and $m + 1$ follow the position of the integer multiple of the laser frequency $m \omega_\text{L}$. To confirm these predictions, we measured the PL spectra of $86$ nm-long gold nanorods terminated by cetyltrimethylammonium bromide (CTAB) with a diameter of $22$ nm and a volume of $3.3 \pm 0.6 \cdot 10^{-23}$~m$^3$.



The nanoparticles were dropcasted from a sufficiently diluted aqueous solution onto a standard microscope cover slip after thorough substrate cleaning by ultrasonication with an alkaline cleaning solution (Hellma Hellmanex, 3\% concentration) and 30~min of UV-ozone treatment to remove residual fluorescence from surface contamination. Excitation by ultrafast laser pulses (Coherent Chameleon Ultra, 80 MHz pulse repetition rate) through a 500~mm-focal-length lens and an oil-immersion, high-numerical-aperture microscope objective (Olympus 60X TIRF, 1.49 numerical aperture) homogeneously illuminated surface regions $20~\mu$m in diameter, allowing for control of the illumination irradiance. Detection of the emitted luminescence from isolated, resolution-limited ($<0.4$~µm) luminescent spots on the sample surface occurred through the same objective using a 680~nm shortpass filter (Semrock) to block the scattered laser radiation. The spectra were dispersed using a reflective grating with 300 grooves per mm in a spectrograph of 300~mm focal length (Princeton Instruments) and detected by a cooled charge coupled device camera (Princeton Instruments Pixis 100B). Integration times between 2~s and 30~s per spectrum were used for the data discussed here. All measurements were conducted at ambient temperatures in air.

\begin{figure}[h]
\centering{\includegraphics[scale=1.2]{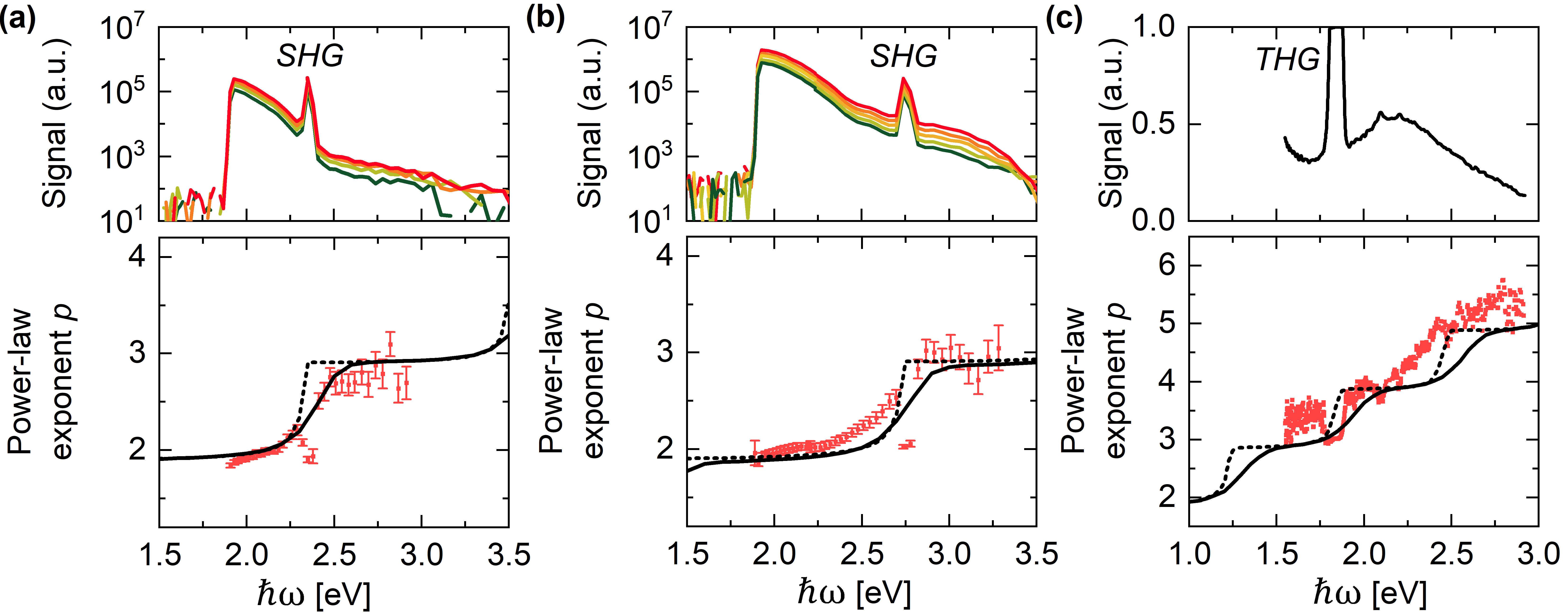}}
\caption{Emission spectra (top panels) and power-law spectra (bottom panels) for 86~nm-long gold nanorods excited at (a) 1.170~eV, 160~fs, $1.2 - 2.2$~kW/cm$^2$ average intensity and (b) at 1.375~eV, 144~fs, $1.8 - 3.1$~kW/cm$^2$. Emission below 1.9~eV is blocked by an optical filter. Panel (c) shows data taken from Ref.~\cite{Knittel_Leitenstorfer_Brida_PRB_2017} for $320$~nm-long gold nanorods excited with 130-fs pulses at a photon energy of 0.61~eV. Solid (dashed) black lines show the power-law exponents in the emission expected from a simplified model neglecting $e-e$ and $e-ph$ scattering for an initial temperature of 500~K (5~K). Coherent nonlinear optical scattering signals corresponding to second-harmonic generation (SHG) and third-harmonic generation (THG) are marked in panels (a)-(c) and have an exponent of 2 and 3, respectively.}
\label{fig:exp}
\end{figure}

Figure~\ref{fig:exp}(a) shows the luminescence spectra for illumination by 160~fs-long laser pulses at $\hbar \omega_{\text{L},0} = 1.170$~eV, below their localized plasmon resonance at 1.62~eV (see Appendix~\ref{app:rods} for the extinction spectrum). Apart from the filter cutoff at 1.91~eV, the spectra cover the whole visible range. The narrow spectral feature is assigned to surface second-harmonic generation (SHG).

Emission spectra were collected for a set of irradiances between 1.2~kW/cm$^2$ and 2.2~kW/cm$^2$, while ensuring the absence of photodegradation by comparing results for upwards and downwards sweeps of the laser power. The power-law coefficient $p$ was extracted as described before by fitting a linear relationship to a double-logarithmic representation of luminescence signal strength and excitation power. Details governing the reproducibility of laser power sweeps and the fit procedure are given in Appendix~\ref{app:power-law-exponent}. At the lowest emitted photon energies, the resulting power-law spectrum in the lower panel of Fig.~\ref{fig:exp}(a) shows a power-law coefficient of 2 similar to the results discussed in~\cite{Lupton_transient_metal_PL_2017}. At close to twice the incident photon energy, the distinct, smooth transition between $p = 2$ and $p = 3$ expected from the theory presented above is indeed found, except for the spectral position of the SHG, for which $p$ drops to 2 as expected. The staircase-like power-law spectrum can be modeled reasonably well even without including electron-electron or electron-phonon scattering, with the shape of the transition region being determined by the average phonon temperature, which based on this analysis is estimated at approximately 500~K. Figure~\ref{fig:exp}(b) shows the same sample region illuminated by 144~fs-long laser pulses at $\hbar \omega_{\text{L},0} = 1.374$~eV, closer to the plasmon resonance, for a similar range of irradiation levels between $I_\text{av} =$ 1.8~kW/cm$^2$ and $I_\text{av} =$ 3.1~kW/cm$^2$. Although approximately one order of magnitude brighter, the luminescence spectra are comparable to the situation before, with the spectral position of the SHG signal shifted towards higher photon energies. The power-law spectra again show a staircase-like transition between $p = 2$ at low photon energies and $p = 3$ at higher photon energies, with the transition region shifted to twice the incident photon energy. This confirms that the observed step-like power-law spectrum is directly linked to the occurrence of step edges in the electron distribution at integer multiples of the laser photon energy above the Fermi level. This signature of the non-thermal nature of the emission at moderately high illumination level is found to be in line with the estimates of the maximal electron temperature reached (see~\cite{Del_Fatti_ultrafast_NLTY_JPCB_2001} or Appendix~\ref{app:rods}), namely, 345~K for the data in Fig.~\ref{fig:exp}(a) and 500~K for the data in Fig.~\ref{fig:exp}(b), as well as an analysis of the effect of temperature on the shape of the step edges. The lower panels of Fig.~\ref{fig:exp} plot the shape of the power-law spectra expected for an electron temperature of 500~K, calculated in the absence of electron-electron or electron-phonon scattering (see Appendix~\ref{app:stepedges}). Indeed, for these low temperatures, the thermal emission still peaks at wavelengths far into the mid infrared, such that the overall emission is dominated by the non-thermal terms.

Higher-order power-law coefficients can only be reached by longer-wavelength laser excitation to avoid the photoionization threshold. Fig.~\ref{fig:exp}(c) shows data extracted from~\cite{Knittel_Leitenstorfer_Brida_PRB_2017} for gold nanorods of $320$~nm length excited by 130~fs laser pulses at $\hbar \omega_{\text{L},0} = 0.61$~eV and 40 MHz pulse repetition rate. The power-law spectrum can be understood reasonably well by our model - the obvious deviations indicate the onset of the thermal emission regime and the increased relevance of electron-electron and electron-phonon scattering over the relevant timescale of the luminescence.

\section{Summary and outlook}\label{sec:summary}
In this work, we have reconciled the different views on the importance of the thermal and non-thermal components in PL from metals following illumination by a short laser pulse. Specifically, we find that in the moderately high and ultrashort illumination limit, the early as well as time-integarted emission is dominantly non-thermal; the behaviour in this regime is thus similar to what happens for emission under CW illumination~\cite{Sivan-Dubi-PL_I}, for which the illumination intensity is relatively weak and the non-equilibrium component in the electron distribution persists permanently. The dependence on the electric field strength in this regime is given by a ``staircase'' of integer powers (see Figs.~\ref{fig:schematic},~\ref{fig:SB_power_law_analysis}(b) and~\ref{fig:exp}), smoothed by the electron temperature. This dependence is rooted in the absorption process. Under these conditions, analysis that relies only on the Planck term (second term in Eq.~(\ref{eq:Gamma_guess}), as e.g., in~\cite{metal_luminescence_Cahill_PNAS}) is valid. However, the emission at the later stages and/or for more intense (e.g.,~\cite{Bouhelier_JOSAB_2021}) pulses gradually attains more and more thermal characteristics. Indeed, under such conditions, the transient effective temperature in the Planck term grows significantly so that the thermal emission shifts from the infra-red to the visible range and becomes stronger. In this case, one should interpret the emission using only the first (Planck) term in Eq.~(\ref{eq:Gamma_guess}) (as, e.g., in~\cite{Greffet_PRX_2018,Bouhelier_JOSAB_2021}) and reference to the emission as thermal light emitted by an object with a time-varying temperature (see~\cite{metal_luminescence_Cahill_PNAS,Lupton_transient_metal_PL,Lupton_transient_metal_PL_2017,Greffet_PRX_2018,Greffet_Adv_Opt_Mat_2019}) is valid as well\footnote{This is similar to the observations regarding the CW PL from a semiconductor (except, possibly for the exceptionally high illumination intensities)~\cite{Sarkar-Un-Sivan-Dubi-NESS-SC}, which are correctly referred to as ``PL by thermalized systems''~\cite{Greffet_PRX_2018,Greffet_Adv_Opt_Mat_2019}.}.

The model we used to explain the PL and to match the experimental data considers single-photon intraband absorption. However, multiphoton interband absorption events may also affect the PL dynamics and the spectrum, especially at high intensities and for high frequency emission. In fact, most earlier studies associated the PL with such interband transitions (see e.g.,~\cite{Mooradian_metal_luminescence,Quidant_modes,Boyd-Shen-luminescence,Hecht_PRL-2019}), even when photons of energies lower than the interband threshold were used; in this case, the emission must indeed originate from {\em simultaneous} 2PA events. Unfortunately, it is very hard to model this effect properly, because the 2PA cross-section is not well characterized and because of the complexity of the d ('valence') bands. In contrast, our model accounts for {\em uncorrelated} photon absorption events within the conduction band, an effect which yields a similar higher-order dependence on the electric field strength associated with the emission spectrum as described above. Notably, uncorrelated multiphoton absorption events are far more likely compared with simultaneous ones, as observed in Ref.~\cite{NUS_2_pulse_PL,Cluzel_PL_2016}; for the same reason, radiative recombination is more likely to be responsible for one-photon emission compared with electronic Raman transitions (see discussion in~\cite{metal_luminescence_Cahill_PNAS,Baffou-aSE-thermometry-review}).

Further studies are required to resolve these issues, as well as to decipher the complex polarization dependence of NPs of more complicated structures under very high excitation intensities~\cite{Giessen_ACS_photonics_2021} and complex size dependence (e.g.,~\cite{Feldman_QY_metal_PL,Tigran-PL,Hecht_PRL-2019,Abajo_DTU_PL}). For these purposes, time-resolved PL spectroscopy (as employed, e.g., in~\cite{koreans_time_resolved_PL_2002,Feldman_QY_metal_PL,Dantus_2015,Suemoto_PRB_2019,Ono,Koyama-2021}) will be of great value and the theoretical framework would need to be expanded beyond the current perturbative approach. Such a formulation would also enable one to determine the quantum yield of the emission process and its parametric dependence. Our work also allows for quantitative interpretation of transient thermometry studies based on the aSE~\cite{Orrit-Caldarola_T_measure_transient}, and associated two-pulse PL experiments (e.g.,~\cite{NUS_2_pulse_PL,Cluzel_PL_2016}) and will be crucial for further studies of other emission processes, such as cathodoluminescence and electroluminescence.

\bigskip

{\bf Acknowledgements.} I.W., I.K. and Y.S. were partially funded by a Lower-Saxony - Israel collaboration grant no. 76251-99-7/20 (ZN 3637) as well as an Israel Science Foundation (ISF) grant (340/2020).

\appendix

\section{Model description}\label{app:model}
The photoluminescence is determined perturbatively from the non-equilibrium electron distribution, which in turn, is determined by solving the Boltzmann equation. This is done using a standard formulation, hence, below, we just describe the approach, point out the novelty in the context of PL studies, and direct the reader to the earlier literature for further details.

The Boltzmann equation employed in the current work incorporates three generic terms~\cite{Dubi-Sivan,Dubi-Sivan-Faraday}. First, the $e-e$ interaction term is responsible for thermalizing the conduction electrons; it is described by the (standard) 2-particle interaction integral following Fermi’s golden rule (see e.g.,~\cite{delFatti_nonequilib_2000,non_eq_model_Rethfeld,vallee_nonequilib_2003,Italians_hot_es,non_eq_model_Rethfeld_con,Ono}), i.e., avoiding the relaxation time approximation. For simplicity, we assume momentum-independent hard-sphere interactions, but verified that the results described below do not change qualitatively if we change the strength of the $e-e$ interactions\footnote{This can be thought of as a way to mimic $e-e$ interactions more accurately, such as the momentum-dependent Thomas-Fermi formulation, etc.}. Second, the $e-ph$ interaction term is responsible for cooling the electron subsystem; it is computed via the deformation potential scattering as in Ref.~\cite{delFatti_nonequilib_2000,Dubi-Sivan}. In this context, for simplicity, the phonon modes are described by the Debye model with a linear frequency dispersion. We further assume that the phonon system is in equilibrium, so that the number of phonons is given by the Bose-Einstein distribution. Additionally, we set $T_{\text{ph}} = 300$ K, and thus, neglect also heat transfer to the environment; accounting for these aspects (e.g., via a TTM) would hardly affect any of the results in this work. Third, the key novelty in our PL formulation concerns the explicit inclusion of the photon-electron interaction, which is based on the formulation described in Refs.~\cite{delFatti_nonequilib_2000,GdA_hot_es,Dubi-Sivan}. Specifically, the photon-electron interaction is weighted by the appropriate combination of the population functions so that the transitions occur between occupied and unoccupied states. As in Refs.~\cite{delFatti_nonequilib_2000,GdA_hot_es}, the temporal profile of the photon-electron interaction term is assumed to be the same as that of the incident pulse. This form enables us to monitor the electron distribution in the early stages of the dynamics.

Finally, it is worth mentioning that the distribution of the occupied states $f$ in the low-electron-energy region is very close to 1. As a result, the distribution of the unoccupied states (denoted by $f_\text{h}$ hereafter) cannot be accurately represented by simply calculating $1-f$ due to machine precision limitations. Therefore, evaluating the emission of high photon frequencies (usually, higher than $2 \hbar \omega_{\text{L},0}$) by substituting $f$ into Eqs.~\eqref{eq:I_e} and~\eqref{eq:rho_J} may lead to inaccurate results. Although this issue could possibly be resolved by increasing the machine's precision, doing so would require a significant amount of computational time and resources. To address this issue without compromising feasibility and accuracy, we also solve the Boltzmann equation for $f_\text{h}$ subject to the initial condition $f_\text{h}(\mathcal{E},t\rightarrow - \infty) = 1 - f(\mathcal{E},t\rightarrow - \infty) = \left(e^{-(\mathcal{E}-\mu)/k_\text{B} T_\text{e,0}} + 1\right)^{-1}$, where $T_{e,0}$ is the initial electron temperature. The Boltzmann equation for $f_\text{h}$ is the same as the one for $f$ but with all interaction terms assigned the opposite sign, such that electron number conservation is guaranteed.

We emphasize that our model accounts for two (or more) photon absorptions from consecutive (and hence, uncorrelated) intraband transition events (but not for the much less likely simultaneous absorption event of two or more photons), in line with the observations in Refs.~\cite{NUS_2_pulse_PL,Cluzel_PL_2016}. It is also worth noting that we do not rely on phenomenological terminology nor any rigorous description of the excitation and decay of quantized plasmons; the good match of our theoretical model with the experimental data implies that such a concept is unnecessary for our purposes.

Unlike the CW case~\cite{Dubi-Sivan,Dubi-Sivan-Faraday}, there is no analytical solution for $f$, which now depends in an implicit way on the details of the illumination. We also avoid defining the electron temperature, although we do refer to an effective (or equivalent~\cite{Italians_hot_es}) value for it that emerges to be the actual electron temperature in the later stages of the dynamics (see, e.g.,~\cite{delFatti_nonequilib_2000,Italians_hot_es}; this effect can be evaluated with the approach in~\cite{Abajo_DTU_PL}). Also note that in this approach there is no assumption on the scaling of the heat capacity with temperature, as this quantity emerges naturally from the rigorous electronic calculations.

\section{Determination of power-law exponents}\label{app:power-law-exponent}
\begin{figure}
    \centering
    \includegraphics[width=11cm]{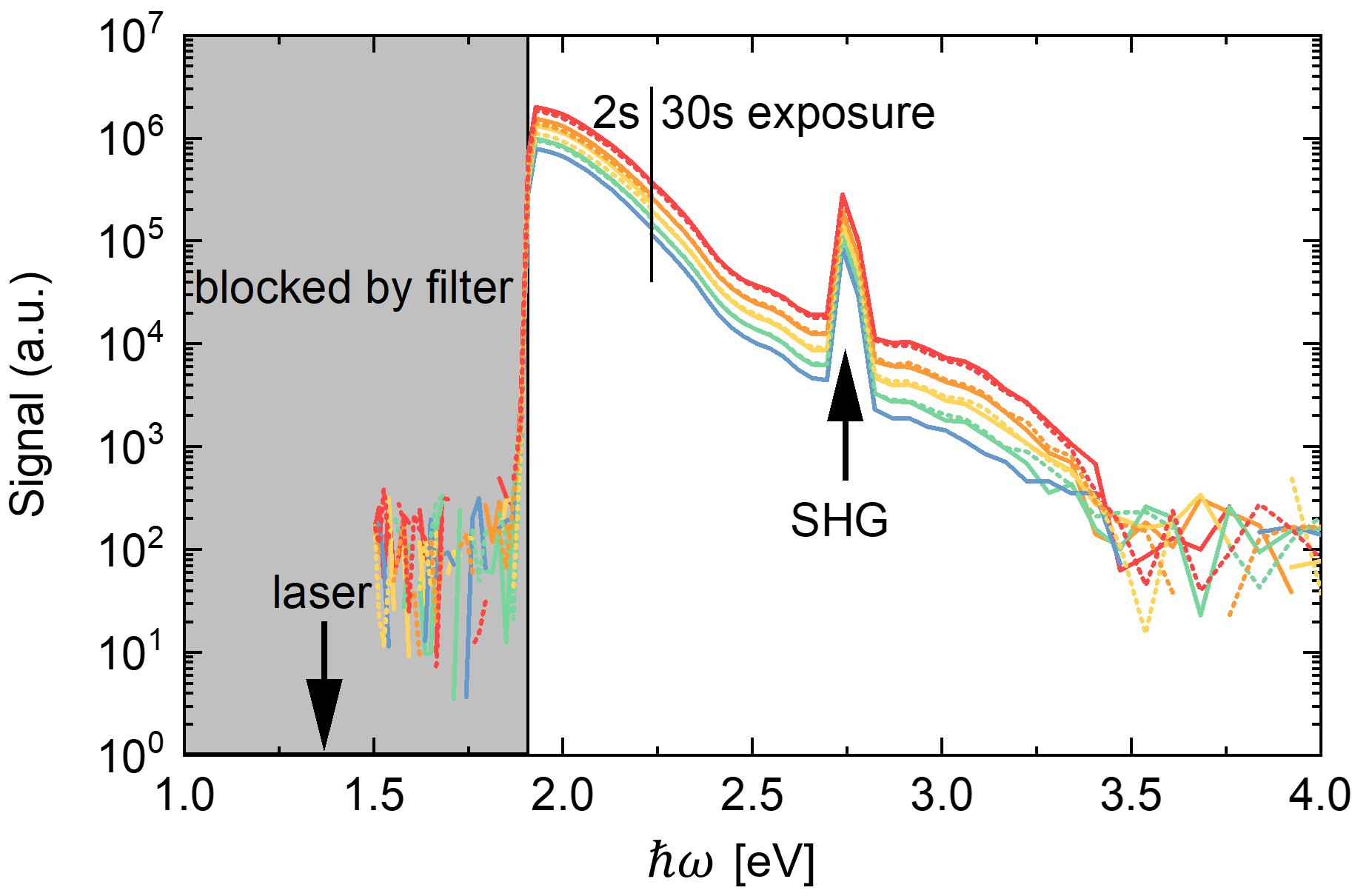}
    \caption{The background-corrected emission spectra resulting for five different levels of irradiance, measured in a bidirectional sweep starting at the highest irradiance. Spectra for the backsweep (increasing irradiance levels) are shown as dashed lines. To fully cover the dynamic range, the emission spectra are stitched from two measurements at either 2~s (below 2.23~eV emission energy) or 30~s (above 2.23~eV) exposure time. Signal data for the 2~s dataset was multiplied by a factor of 15 to correct for the reduced exposure time. Spectral data beyond 3.3~eV is too noisy for further evaluation. }
    \label{fig:SpectraExampleDetails}
\end{figure}

\begin{figure}
    \centering
    \includegraphics[width=\linewidth]{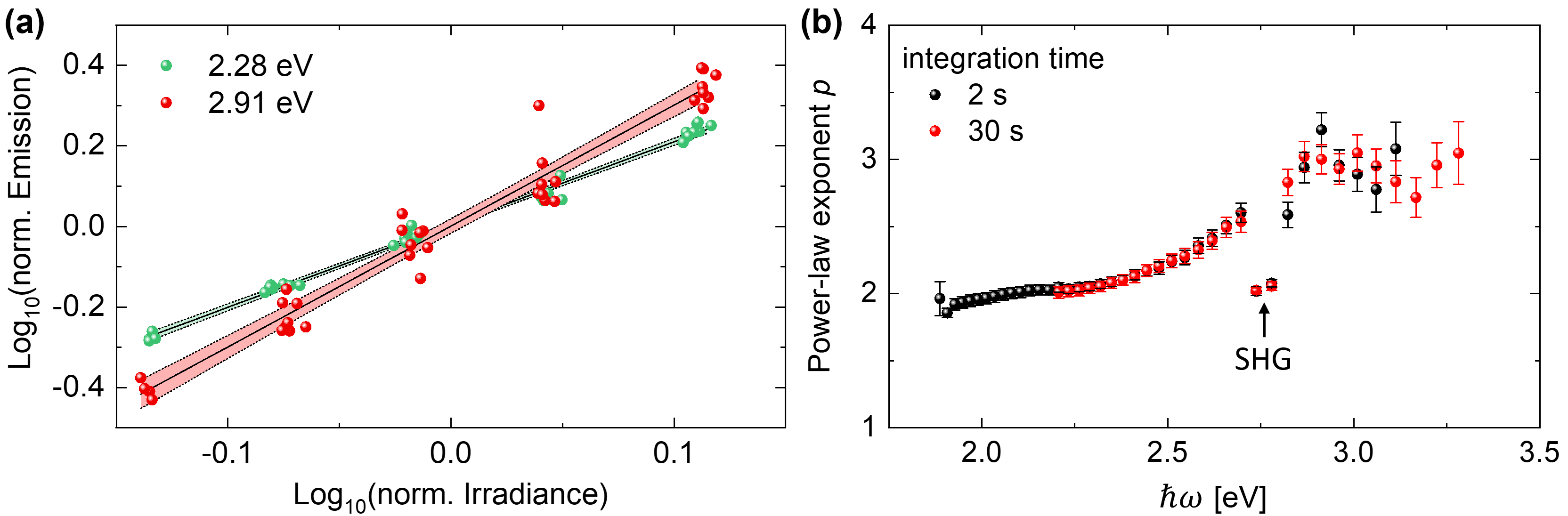}
    \caption{Panel (a) shows the emission measured at photon energies of 2.28~eV (green) and 2.91~eV (red) for five irradiance levels, based on 5 consecutive bidirectional sweeps of irradiance resulting in a total of 36 data points at each photon energy. Data is plotted on normalized logarithmic scales. The results of linear fits (and 95\% confidence limits) are shown by solid (dashed) lines. Panel (b) shows the calculated power-law exponents for two independent measurements at 2~s (black) and 30~s (red) exposure time per spectrum.     }
    \label{fig:PowerLawFitDetails}
\end{figure}

For each power-law exponent fit, individual emission spectra are collected for a set of 5 irradiance levels. For a single bidirectional sweep of the irradiance, a total of 9 spectra are collected, starting and ending at the highest irradiance level. Fig.~\ref{fig:SpectraExampleDetails} shows details for 1.375~eV excitation. Raw data is collected in two consecutive measurement runs using either 2~s or 30~s exposure time of the spectrometer camera, in order to fully cover the dynamic range of the emission spectrum. For emitted photon energies below (above) 2.23~eV, data is shown from the 2~s (30~s) exposures. The figure shows the result of an individual sweep of irradiance levels, with the spectra belonging to the backsweep (towards higher values) of excitation power plotted as dashed lines. The emission spectra are reproducible over multiple consecutive sweeps. Fig.~\ref{fig:PowerLawFitDetails}(a) shows the measured emission intensities at photon energies of 2.28~eV and 2.91~eV for all 5 irradiance levels chosen and all 5 bidirectional sweeps. Irradiance values are those measured directly in the setup and thus randomly scatter around their nominal values. The emission shows random scatter, but no systematic drift as a function of measurement time. The slope of linear fits on the double-logarithmic scale is used to determine the power-law exponents $p$ at a given photon energy. Fig.~\ref{fig:PowerLawFitDetails}(b) shows the results of the consecutive measurement of the power-law exponents based on 2~s and 30~s spectrometer integration time, each measurement consisting of a total of 36 individual spectra. Results are only shown for those spectral regions for which the individual spectra are unaffected by either too long or too short camera exposure. The error bars indicate the standard error of the slope of the linear fit in the double-logarithmic representation shown in Fig.~\ref{fig:PowerLawFitDetails}(a), as reported by the \emph{Mathematica} software package (Wolfram Research). Within the stated error bars, the two measurements yield the same power-law exponent spectrum in Fig.~\ref{fig:PowerLawFitDetails}(b). Note that in the region of SHG, the power-law exponent tends towards 2.

\section{The shape of power-law coefficient spectra at moderate irradiances}\label{app:stepedges}
\begin{figure}[h]
    \centering
    \includegraphics[scale=1.3]{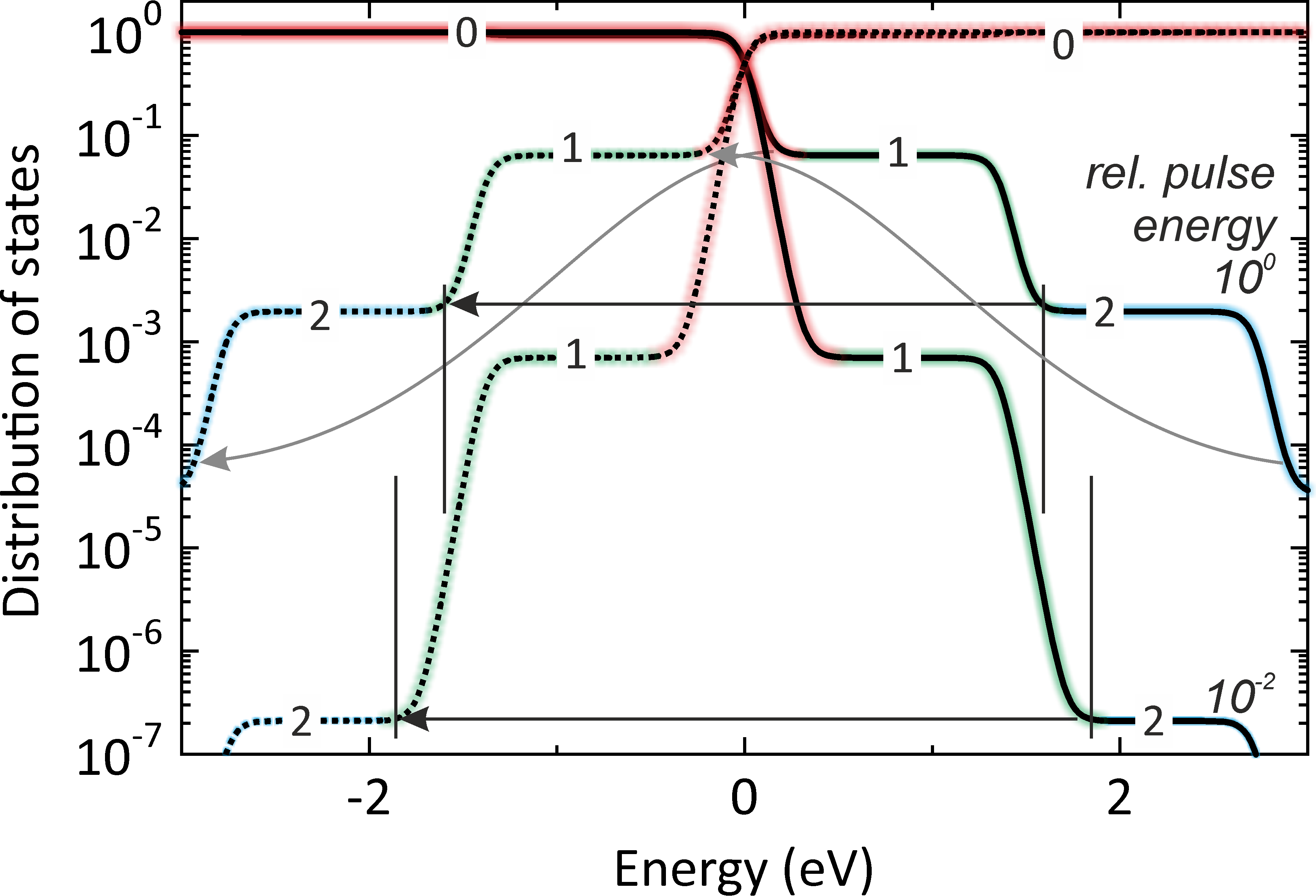}
    \caption{Schematic electron energy distribution relative to the Fermi edge after excitation by 1.375~eV photons. Occupied states (solid lines) and unoccupied states (dashed lines) are shown for two levels of the excitation pulse energy. The effective nonlinear interaction order between photoexcitation and electron occupation associated with each step-like structure is marked by integer digits and red (0), green (1) and blue (2) colors. The maximum photon energy for emission associated with each effective nonlinearity depends on the excitation strength, as marked by the curved black arrows. Note that the other transitions contribute equally (gray arrows).}
    \label{fig:ElecDistSchematic}
\end{figure}

\begin{figure}[h]
    \centering{\includegraphics[scale=1.25]{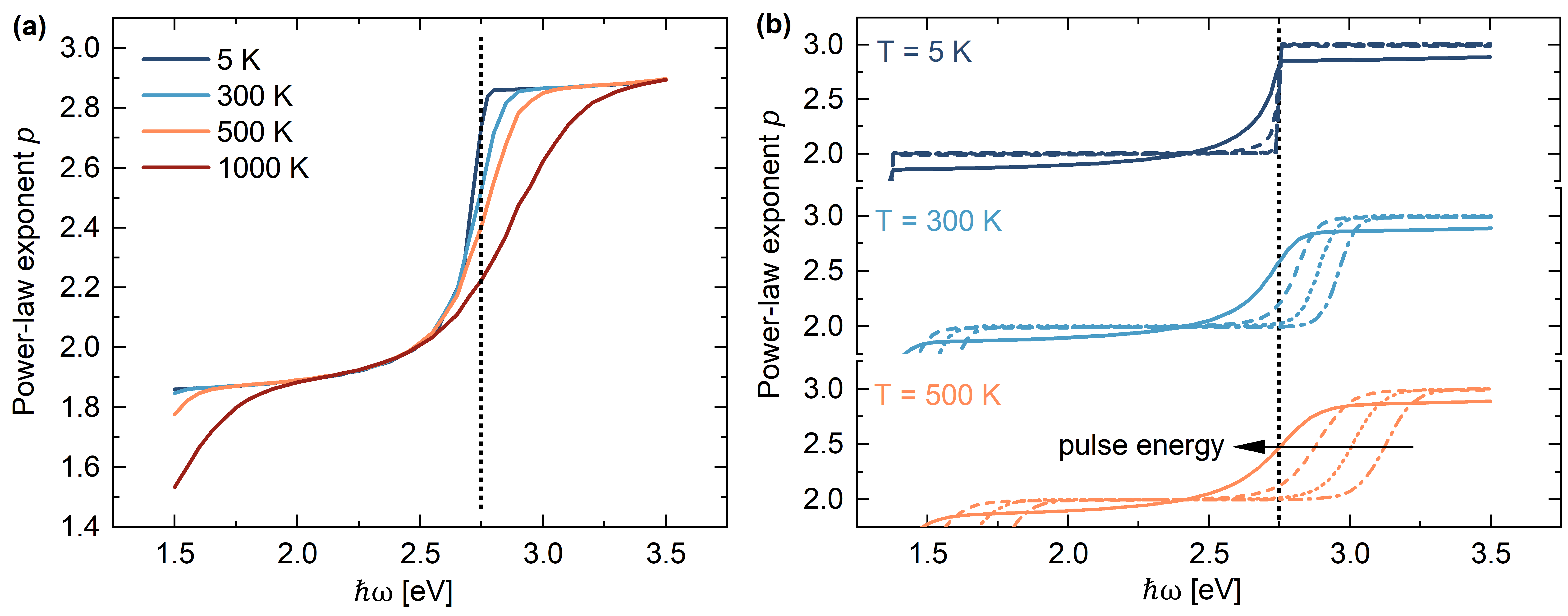}}
    \caption{Variation of the shape of step edges in the power-law exponents of the emission spectrum for excitation at 1.375~eV photon energy. Panel (a) shows the effect of the initial electron temperature before taking electron-electron and electron-phonon scattering into account, in the regime of moderately high laser pulse energies. Panel (b) shows the effect of changing the excitation pulse energy at three different temperatures. The dotted line marks a photon energy of 2.75~eV, twice the energy of the excitation photons. Pulse energies reduced by factors of $10^{-1}$, $10^{-2}$, and $10^{-3}$ are shown as dashed, dotted, and dash-dotted lines.}
    \label{fig:stepshape}
\end{figure}

For moderate excitation pulse energies, the shape of the power-law exponent spectrum is determined to a large degree by the electron distribution present before the arrival of the laser pulse. Fig.~\ref{fig:ElecDistSchematic} shows the electron distribution for laser excitation at 1.375~eV together with typical transitions contributing to the PL emission. Before accounting for electron-electron or electron-phonon scattering, photoexcitation of the electron distribution results in characteristic ``shelves'' of the density of states (estimated as constant, for simplicity), each being associated with an effective integer nonlinearity. These shelves are connected by Fermi-like transitions. For a chosen emitted photon energy, the characteristic interaction orders for the occupied and the unoccupied states add up to the associated total effective nonlinearity, or power-law coefficient. To determine the power-law exponent spectra, the emission is calculated in the range of 80\% to 125\% of a given reference excitation pulse energy and fitted linearly on a double-logarithmic scale. The results are plotted against the emitted photon energy in Fig.~\ref{fig:stepshape}. An increase of the initial electron temperature softens the transition regions between individual plateaus in the staircase-like power-law spectrum. For the lower excitation pulse energies, increased temperatures induce pronounced shifts of the transition region towards higher photon energies. This effect is caused by the increased occupation in the high-energy tails of the Fermi distributions, cf. Fig.~\ref{fig:ElecDistSchematic}. This shift can be used as an estimate of the effect of heat accumulation caused by the pulsed excitation at high repetition rates, as the initial electron temperature is equal to the phonon temperature.

\section{Dependence of power-law coefficient spectra on excitation intensity}\label{app:simulatedpowerlawspectra}
\begin{figure}[h]
    \centering{\includegraphics[width=\linewidth]{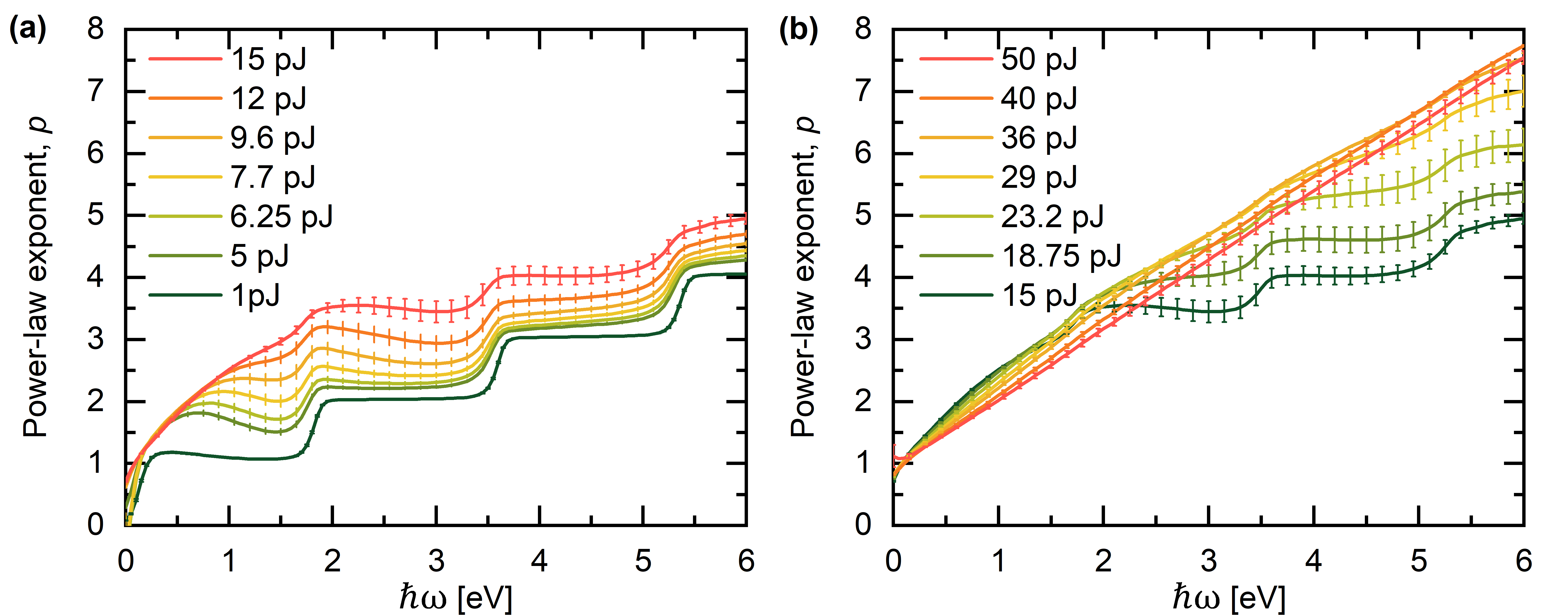}}
    \caption{Power-law exponent spectra calculated from simulated electron dynamics for a range of excitation intensities. For each line, three simulations at 80\%, 100\% and 125\% of the indicated excitation intensity were used. Error bars indicate the degree of divergence from a pure power-law excitation dependence.}
    \label{fig:simulatedpowerlawspectra}
\end{figure}

A gradual increase of excitation intensity causes an evolution of power-law coefficient spectra from a staircase-like appearance to a linear functionality. Fig.~\ref{fig:simulatedpowerlawspectra} shows spectra calculated from simulated electron dynamics over a wide range of excitation intensities, with the simulation parameters being the same as those used for Fig.~\ref{fig:SB_power_law_analysis}. For each indicated excitation intensity, the total emission at each photon energy for three simulated electron dynamics at 80\%, 100\% and 125\% of the nominal excitation value was linearly fitted on the double-logarithmic scale, the same procedure as applied to experimental data. Panel (a) shows the evolution in the moderate-to-high excitation intensity regime between 1~pJ and 15~pJ, while panel (b) covers the high-intensity regime between 15~pJ and 50~pJ.

A few notable observations can be made based on this data. First, power-law coefficients at lower photon energies converge towards the thermal, linear functionality at lower intensities than those at higher photon energies. This agrees with the heuristic interpretation, namely, that as the illumination intensity grows, the thermal contribution becomes stronger and shifts from the infrared into the visible spectral range. Second, the errors of fitting the excitation intensity-dependence by a power-law are largest in the regime of intermediate intensities and high photon energies. This indicates that in these regions, power-law functionalities are a poor approximation of the excitation dependence. Third, while in the moderate-to-high excitation intensity regime the power-law in general shows a monotonic increase of $p$ coefficients with intensity, this is no longer true in the high-excitation-intensity regime. In this regime the intensity-dependence indeed reverses and one finds a lowering of $p$ coefficients with increase of excitation intensity. This is in line with the predictions from the analysis of fully-thermalized Fermi-Dirac distributions~\cite{Lupton_transient_metal_PL}, and is mostly an expression of the gradual shape change of the Fermi-Dirac distribution at higher electron temperatures.

\section{Characteristics of the gold nanorod sample}\label{app:rods}
\begin{figure}
    \centering{\includegraphics[scale=1.16]{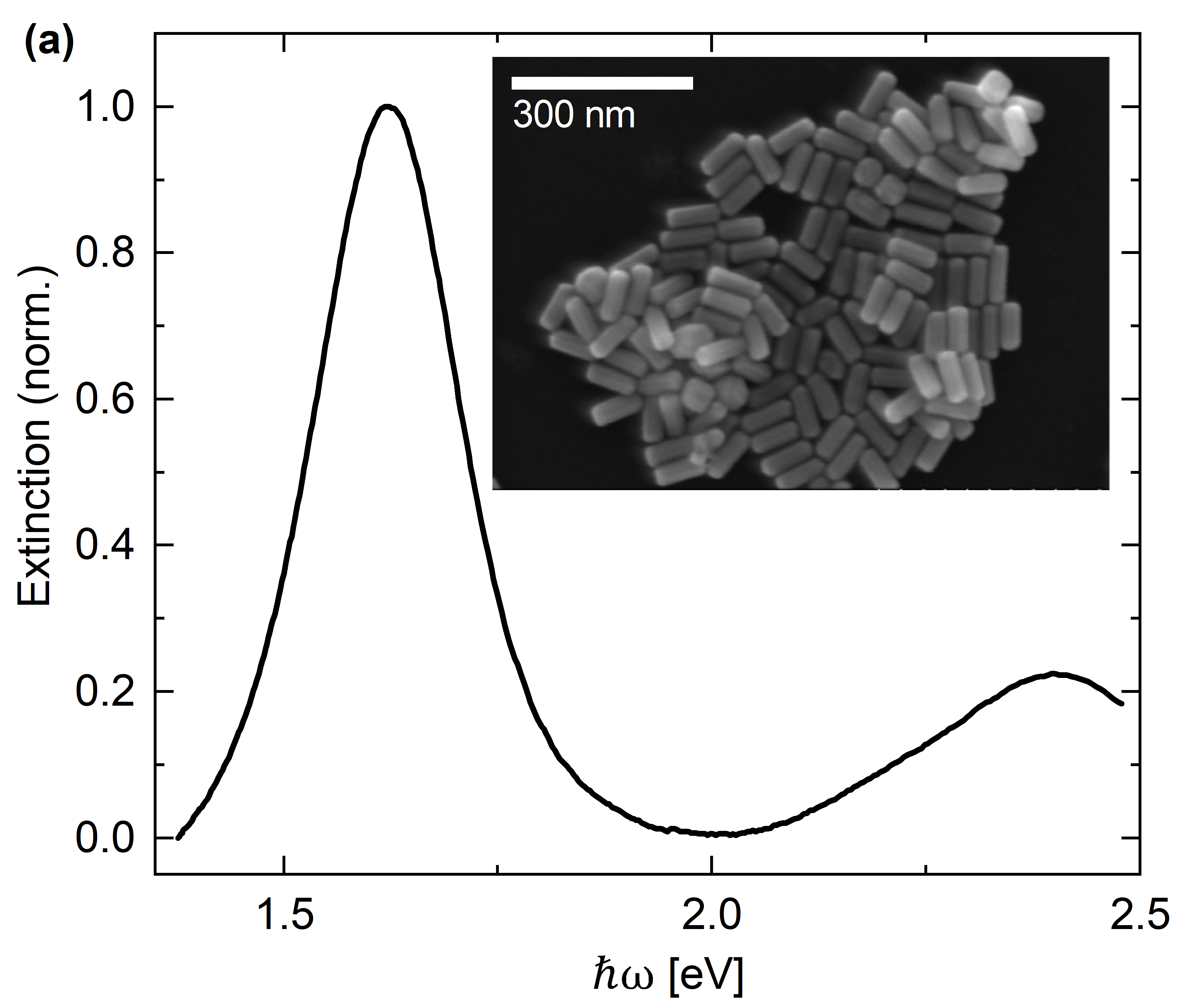}\hfill\includegraphics[scale=0.43]{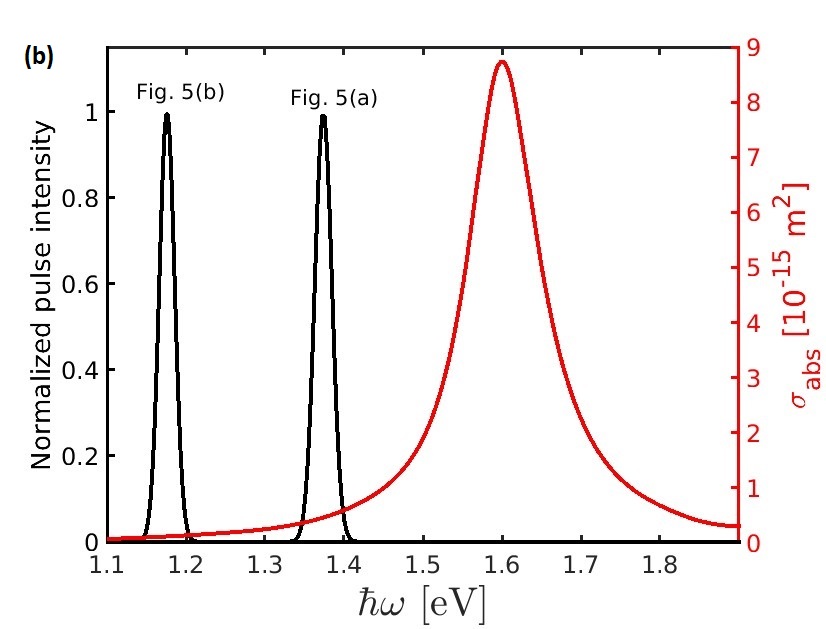}}
    \caption{Panel (a) shows the normalized extinction spectrum for the CTAB-terminated gold nanorods. The inset shows a scanning electron micrograph of the particles. Panel (b) shows the calculated absorption cross-section as a function of wavelength, together with the normalized laser spectra for the wavelengths used in Fig.~\ref{fig:exp}~(a)-(b).}
    \label{fig:nanorods}
\end{figure}

Gold nanorods terminated by cetyltrimethylammonium bromide (CTAB) were synthesized as described in~\cite{size_effect_PL_Xiamen}. Based on a characterization by scanning electron microscopy, the rod length was $86\pm6$~nm, and the diameter $22\pm 2$~nm. The extinction spectrum and an SEM image is shown in Fig.~\ref{fig:nanorods}(a). Samples were fabricated on thoroughly cleaned microscope cover slips by dropcasting. Fig.~\ref{fig:nanorods}(b) shows the absorption cross-section $\sigma_\text{abs}(\omega)$ of the rods, along with the spectra of two pulses used to generate the data in Fig.~\ref{fig:exp}~(a)-(b).

Following~\cite{Del_Fatti_ultrafast_NLTY_JPCB_2001}, it can be shown that the maximal electron temperature can be approximated by
\begin{equation}\label{eq:max_Te}
\text{max}(T_\text{e}) \approx \sqrt{T_\text{env}^2 + 2 Q / \gamma},
\end{equation}
where $Q = \sigma_\text{abs}(\omega) I_0 \tau_\text{L} / V$ is the power absorbed by the nanorod following illumination by a single pulse (see Section~\ref{subsec:numerics}), $V$ the rod volume, and $\gamma = 67$~J/m$^3$K$^2$ the proportionality constant between the electron heat capacity and the electron temperature (i.e., $C_\text{e} = \gamma T_\text{e}$~\cite{Ashcroft-Mermin}).

As this relation neglects the non-thermal stage of the dynamics, as well as energy transfer from the electrons to the phonons, it provides no more than a reasonable estimate to the actual heat 
dynamics. Nevertheless, for the current purposes, it enables an estimate of the wavelength of maximal thermal emission via Wien's displacement law, $\lambda_\text{peak} = c / \nu_{max}$ where $\nu_{max} = 5.879 \cdot 10^{10} \text{max}(T_\text{e})$.


\end{document}